\documentclass{aa}
\usepackage{graphicx}
\def\ltsima{$\; \buildrel < \over \sim \;$}    % Use in text mode
\def\lesssim{\lower.5ex\hbox{\ltsima}}           % Use in math mode
\def\gtsima{$\; \buildrel > \over \sim \;$}    % Use in text mode
\def\gtrsim{\lower.5ex\hbox{\gtsima}}           % Use in math mode

\def\gamef{{\gamma_{\rm e}}}
\def\Ma{M_{\rm a}}
\def\VS{V\'azquez-Semadeni}
\def\VV{V}
\input{psfig.sty}
\begin{document}
   \title{Magnetic Pressure-Density Correlation in Compressible MHD
Turbulence}

%   \subtitle{}

   \author{Thierry Passot
          \inst{1}
          \and
          Enrique V\'azquez-Semadeni\inst{2}
%\fnmsep
          }

%   \offprints{}

   \institute{CNRS, Observatoire de la C\^ote d'Azur, B.P. 4229,
06304, Nice, C\'edex 4, France\\
%              \email{passot@obs-nice.fr}
         \and
             Instituto de Astronom\'\i a, UNAM, Unidad Morelia, Apdo.\
Postal 7-32, Morelia, Michoac\'an, 58089, M\'exico\\
%             \email{e.vazquez@astrosmo.unam.mx}
             }

   \date{}

   \abstract{
We discuss, both analytically and numerically, the behavior of magnetic
pressure and density fluctuations in strongly turbulent isothermal
magnetohydrodynamic (MHD) flows in ``1+2/3'' dimensions, or ``slab''
geometry. We first consider ``simple'' MHD waves, which are the
nonlinear analogue of regular MHD waves, and have the same three modes,
slow, fast and Alfv\'en. These allow us to write equations for the
magnetic field strength $B$ as a function of density $\rho$ for the slow
and fast modes, showing that the two have different asymptotic
dependences of the magnetic pressure $B^2$ vs.\ $\rho$. For the slow
mode, $B^2 \simeq c_1 - c_2 \rho$, while for the fast mode, $B^2 \simeq
\rho^2$. We also perform a perturbative analysis to investigate Alfv\'en
wave pressure, recovering the results of McKee and Zweibel that $B^2
\simeq \rho^\gamma_e$, with $\gamma_e \simeq 2$ at large $M_a$, $\gamma_e
\simeq 3/2$ at moderate $M_a$ and long wavelengths, and $\gamma_e \simeq
1/2$ at low $M_a$.  This wide variety of behaviors implies that a single
polytropic description of magnetic pressure is not possible in general,
since the relation between magnetic pressure $B^2$ and density is not
unique, but instead depends on which mode dominates the density
fluctuation production. This in turn depends on the angle $\theta$
between the magnetic field and the direction of wave propagation and on
the Alfv\'enic Mach number $M_a$. Typically, at small $M_a$, the slow
mode dominates, and $B$ is {\it anti}correlated with $\rho$. At large
$M_a$, both modes contribute to density fluctuation production, and the
magnetic pressure decorrelates from density, exhibiting a large scatter,
which however decreases towards higher densities. In this case, the
magnetic ``pressure'' does not act as a restoring force, but rather as a
random forcing. These results have implications on the probability
density function (PDF) of mass density. The unsystematic behavior of the
magnetic pressure causes the PDF to maintain the lognormal shape
corresponding to non-magnetic isothermal turbulence, except in cases
when the slow mode dominates, in which the PDF develops an excess at low
densities because the magnetic ``random forcing'' becomes density
dependent. Our results are consistent with the low values and apparent
lack of correlation between the magnetic field strength and density in
surveys of the lower-density molecular gas, and also with the
recorrelation apparently seen at higher densities, if the Alfv\'enic
Mach number is relatively large there.

%-- Density peaks <--> magnetic minima
%-- No polytropic description
}

   
\authorrunning{Passot \& V\'azquez-Semadeni}

\titlerunning{Density-magnetic pressure correlation}

   \maketitle
%
%________________________________________________________________

\section{Introduction}

The cold molecular gas in our Galaxy is generally believed to be
magnetized and turbulent (see, e.g., the reviews by Dickman 1985;
Scalo 1987; Heiles et al.\ 1993; V\'azquez-Semadeni et al.\
2000). However, at present, the actual strength of the magnetic field
in molecular clouds and, as a consequence, its dynamical importance,
are still a matter of debate, with opinions ranging from considering
it crucial (e.g., Crutcher 1999) to moderate (e.g., Padoan \& Nordlund
1999), although in general, the observational evidence so far appears
inconclusive (e.g., Bourke et al.\ 2001; Crutcher, Heiles \& Troland
2002). Therefore, it is important to continue gathering information both 
observationally and theoretically on the distribution and strength of
the magnetic field in the ISM and, in particular, molecular clouds.

In general, magnetic fields and turbulence are believed to provide
support against the self-gravity of the clouds, which typically
have masses much larger than their thermal Jeans mass. In this context,
the pressure provided by magneto-hydrodynamic (MHD) motions in the
clouds has been investigated in a variety of scenarios. Since
observations often suggest that there is near equipartition between
turbulent and magnetic energies in molecular clouds (e.g., Myers \&
Goodman 1988; Crutcher 1999), the motions are often considered to be
MHD waves (Arons \& Max 1975). Until recently, Alfv\'en waves were favored,
as these were expected to be less dissipative than their slow and fast
counterparts, but recent numerical simulations (Mac Low et al.\ 1998;
Stone, Ostriker \& Gammie 1998; Padoan \& Nordlund 1999) have shown
that supersonic MHD turbulence decays almost as rapidly as purely
hydrodynamic turbulence because of strong coupling between the MHD
modes, and so the rationale for the preponderance of the Alfv\'en mode
appears less clear now (but see Cho, Lazarian \& Vishniac 2002 for an
argument against fast decay). It is thus necessary to investigate all
MHD modes as a source of pressure in a general context in fully
turbulent regimes.

%Another consequence of those dissipation studies, when considered
%together with the fact that all large enough clouds exhibit supersonic
%motions (Larson 1981; see also Blitz 1993 for a more recent review),
%is that the turbulence in molecular clouds must somehow be constantly
%replenished, either 
%from their internal star-forming activity (e.g., Allen \& Shu 1999),
%from external passing shock fronts (Kornreich \& Scalo
%2000), or from the formation of the clouds themselves, through
%bending-mode instabilities (\VS, Ballesteros-Paredes \& Klessen
%2002). Another perspective is that molecular clouds may have much 
%shorter lifetimes than previously assumed (Ballesteros-Paredes,
%Hartmann \& V\'azquez-Semadeni 1999). 

The functional form that the magnetic pressure may take as a function of 
density is also important for the statistics of the density fluctuations,
as it was shown by Passot \& V\'azquez-Semadeni (1998; see also
Nordlund \& Padoan 1999) that for flows in which the pressure $P$ exhibits
an effective {\it polytropic} behavior, such that it 
scales with density $\rho$ as $P \sim \rho^\gamef$, with
$\gamef$ the effective polytropic exponent, the shape of the density
probability density function (PDF) depends on the specific value of
$\gamef$. The PDF takes a lognormal form for $\gamef=1$, but it
develops an asymptotic power-law tail at low (high) densities for
$\gamef>1$ ($\gamef<1$). Thus, the value of $\gamef$ produced by the
various sources of pressure is important for the density statistics and
the overall dynamics of the flow.

In an analysis of the pressure produced by Alfv\'en waves, McKee \&
Zweibel (1995, hereafter MZ95) concluded, based on an analysis by Dewar
(1970), that it is isotropic, and that, for flows 
undergoing slow compressions, $\gamef=3/2$, while for strong shocks,
$\gamef=2$. \VS,
Cant\'o \& Lizano (1998) studied the evolution of the velocity
dispersion $\sigma$ in gravitationally collapsing flows, as a means of
determining the density dependence of the ``turbulent pressure''
$P_{\rm turb}$, defined in such a way that $\sigma^2 \equiv dP_{\rm
turb}/d\rho$. They found that the for slowly collapsing magnetized
cases, $\gamef \sim 3/2$, while for rapid collapse, $\gamef\sim 2$.

More recent numerical work has directly plotted the magnetic strength
$B$ or the magnetic pressure
$B^2$ in three-dimensional numerical simulations of both isothermal
(Gammie \& Ostriker 1996; Padoan \& Nordlund 1999; Ostriker, Stone \&
Gammie 2001) and non-isothermal flows (Kim, Balsara \& Mac Low 2001;
MacLow et al.\ 2001). The isothermal simulations 
should be reasonable models of molecular clouds, while the
non-isothermal ones are cast as models of the ISM at large. Gammie \&
Ostriker (1996) report an almost constant (on average) magnetic field
strength as a function of density in non-self-gravitating runs. Padoan
\& Nordlund's (1999) plot of $B$ vs.\ $\rho$ shows a 
scatter of over two dex in the low density range for large values of the
Alfv\'enic Mach number, with the scatter decreasing towards larger
densities. On the other hand, there is less scatter at low Alfv\'enic
Mach numbers ($\Ma$), but it is more uniform throughout the density
range. Those authors note that the upper envelope of their $B$
distributions closely matches the observational scaling $B \sim
\rho^{1/2}$ found in cases when the magnetic field is detected (e.g.,
Crutcher 1999; Crutcher et al.\ 2002), although they remark that in a
large number of cases only 
upper limits or complete non-detections are obtained. Such scaling also
corresponds to the high-$B$ tail of the magnetic strength distribution
in the high-$\Ma$ simulation by Padoan \& Nordlund. It is worthwhile to
additionally note that the 
bulk of the points in their $B$ distribution does not follow a simple
power-law scaling with $\rho$, but instead appears to curve up, being
first nearly constant with $\rho$ and then starting to rise.
Ostriker et al.\ (2001) find similar
results, except that these authors only attempt to fit the high-density
part of the $B$ distribution to observations, finding slopes between
0.3 and 0.5, which again are not too different from the observed
scaling. 

In non-isothermal simulations, Mac Low et al.\ (2001) have plotted the
magnetic pressure vs.\ the thermal pressure, finding again a large
scatter of up to 6 orders of magnitude in their supernova-driven
simulations, while Kim et al.\ (2001) show plots of $B$  vs.\ $\rho$,
finding slopes of $\sim 0.4$, although with scatter of one order of
magnitude at high density, and two orders of magnitude at lower
densities, the plot actually appearing more like two segments with very
different slopes. 
Finally, in simulations of thermal condensation triggered by strong
compressions in the presence of a uniform magnetic field, which are in
several aspects similar to gravitational collapse, Hennebelle \&
P\'erault (2000) found that the magnetic field does not necessarily
increase together with the density.

Recent theoretical work (e.g., Lithwick \& Goldreich 2001; Cho et al.\
2002) has not specifically addressed the issue of the magnetic
pressure-density correlation, as it has focused mainly on the spectral
properties of moderately compressible MHD turbulence, as a consequence
of mode coupling. Nevertheless, Cho et al.\ (2002) mention in passing
that their results lead them to {\it not} expect a significant correlation
between the magnetic pressure and the density.

In an attempt to understand the physics underlying the above
experimental and observational results, in this paper we study the
dependence of magnetic pressure with density as a consequence of
nonlinear MHD wave propagation in isothermal MHD turbulence in
``1+2/3'' dimensions, also referred to as a ``slab'' geometry. This
choice allows us to isolate the relative importance and effects of the
relative orientation of the magnetic field and wave propagation
directions. We first consider the problem analytically, finding asymptotic
relations between the density, magnetic field and velocity field
fluctuations, which show the effectiveness of the slow and fast modes as 
sources of density fluctuations under a variety of conditions (\S
\ref{sec:simple}). We then discuss Alfv\'en wave pressure using a
perturbation analysis, comparing with the results of MZ95
(\S \ref{sec:alfven}). These results are then tested by means of
numerical simulations in the same geometry (\S \ref{sec:num_sim}),
which support the scenario derived from the analysis of the
equations. Finally, in \S \ref{sec:conclusions} we present a summary and
some discussion of our results (\S \ref{sec:summary}), including
implications for the interpretation of observational results (\S
\ref{sec:implications}).

%__________________________________________________________________

\section{Properties of simple waves} \label{sec:simple}

The equations governing, in the magnetohydrodynamic (MHD) limit,
the one-dimensional motions of a plasma permeated by a
uniform magnetic field ${\bf B_0}$ in a slab geometry read

\begin{eqnarray}
&&{\partial\rho\over\partial t} +{\partial(\rho u) \over\partial x}=0
\label{eq:mhd1d1}\\
&&{\partial u\over\partial t}+u{\partial u\over\partial x}
=-{1\over\rho} {\partial\over \partial x} \left ( {\rho\over
M_s^2}+{|b|^2\over 2 M_a^2} \right ) +f_x\label{eq:mhd1d2}\\
&&{\partial v\over\partial t}+u{\partial v\over\partial x} =
{b_x\over M_a^2\rho}{\partial b\over \partial x} +f\label{eq:mhd1d3}\\
&&{\partial b\over\partial t}+ {\partial\over \partial
x}(ub)=b_x{\partial v\over \partial x},
\label{eq:mhd1d4}
\end{eqnarray}\par

\noindent where $x$ is the direction of propagation,
nondimensionalized by a typical length $L$. The velocity components $u$
(along the $x$ axis) and $v=v_y+iv_z$ (along the $y$ and $z$ axes)
are normalized by a velocity unit 
$U_0$, the mass density $\rho$ by a reference density $\rho_0$, and
the magnetic field components $b_x=\cos \theta$ and
$b=b_y+ib_z$ by  $B_0$. Note that $b_x$
is constant and that $b_z(t=0)=\sin \theta$, where
$\theta$ is the angle between the direction of propagation and that of
the initially unperturbed ambient magnetic field. Time $t$ is
measured in units of $L/U_0$. An isothermal equation of state is
assumed and we denote by $c_s$ the constant sound speed.
Two non-dimensional numbers can be defined, the sonic Mach number
$M_s={U_0/c_s}$ and the Alfv\'enic Mach number $M_a={U_0/v_a}$,
where $v_a={B_0/(4\pi \rho_0)^{1/2}}$ is the Alfv\'en speed of
the unperturbed system.
The plasma beta is here defined by $\beta={M_a^2\over
M_s^2}$. The above system of equations also contains a driving in the
form of random accelerations $f_x$ (acting on the component $u$  of
the velocity) and $f=f_y+if_z$, (acting on the components  $v$).

In this section we shall derive some properties of the so-called simple MHD
waves (see, e.g., Landau \& Lifshitz 1987, \S 101) and thus we assume
$f_x=f=0$. In the case where the basic state
is perturbed by infinitesimal disturbances, the solutions are
superpositions of linear plane travelling waves. These plane waves
are monochromatic and their profile does not change in time,
with all quantities only depending on the combination $x\pm Ct$. The
propagation velocity $C$ is a constant that identifies with one of
the three possible roots of the dispersion relation, namely the
slow, Alfv\'en or fast velocity. For a particular plane wave
solution each perturbed quantity can be expressed as a function of a
chosen one, for example the density $\rho$.

In the case of finite-size perturbations these relations do not
hold, but it is nevertheless possible to look for particular
solutions that have the property that all quantities are only
functions of any single one of them, as in the linear theory. These
particular solutions, which generalize the linear plane wave
solutions to the case of finite disturbances, are called simple
waves.

Following Landau and Lifshitz (1987, \S 105; see also Mann 1995), we
recall here how to derive 
the relevant equations that determine simple wave profiles.
The principle is illustrated on the equation for mass conservation,
that can be rewritten in the form
\begin{equation}
{\partial (\rho,x)\over \partial (t,x)}-{\partial (\rho u,t)\over
\partial (t,x)} =0,
\end{equation}
after transforming the partial derivatives into Jacobians.
If one now chooses $(t,\rho)$ as new independent variables, one has to
multiply the above equation by ${\partial (t,x)\over \partial
(t,\rho)}$. Assuming that all dependent variables only depend on
$\rho$, and writing $U={dx\over dt}$ one gets, after expanding the Jacobians,
\begin{equation}
-U+\rho {du\over d\rho}+u=0 ,
\end{equation}
or, more generally,
\begin{equation}
-\VV d\rho +\rho du=0 ,\label{eq:simple1}
\end{equation}
where $\VV=U-u$ denotes the wave speed.
Each field is thus a function $F(x-(u+\VV)t)$ where $u$ and
$\VV$ are functions of e.g. $\rho$. Each point of the wave
is traveling with its own velocity, leaving the possibility for
wave steepening and the subsequent formation of a discontinuity.

Using the same procedure, equations
(\ref{eq:mhd1d2})-(\ref{eq:mhd1d4}) read
\begin{eqnarray}
&&-\VV\rho du+{1\over M_s^2}d\rho+ {1\over M_a^2}d{|b|^2\over
2}=0 \label{eq:simple2}\\
&& -\VV\rho dv-{b_x\over M_a^2}db=0\label{eq:simple3}\\
&& -\VV db+bdu-b_xdv=0.\label{eq:simple4}
\end{eqnarray}

The system
(\ref{eq:simple1})-(\ref{eq:simple4}) has non-trivial solutions if
the wave speed $\VV$ is given by
\begin{equation}
\VV_{\pm}^2={1\over 2M_a^2 \rho}\left ({B^2}+{\beta \rho
} \right ) \left ( 1\pm \sqrt { 1-{4\beta b_x^2\rho \over
(B^2+\beta \rho)^2}} \right ) \label{eq:fs-speeds}
\end{equation}
or if it equals the Alfv\'en speed $\VV_A=\pm {b_x\over M_a
\rho^{1/2}}$. Recall that we denote by $B^2=b_x^2+|b|^2$ the total
magnetic intensity.

The latter root is associated with the circularly polarized Alfv\'en
simple wave, which is non-compressive and has $|b|^2 = cst.$
The solutions (\ref{eq:fs-speeds}) are associated with fast
and slow simple waves. The speeds of the linear fast and slow
magnetosonic waves are recovered by taking  $\rho$=1 and $b=i\sin \theta$.

We are thus led to the following system of equations for the wave
profiles
\begin{eqnarray}
&&{du\over d\rho}={\VV\over\rho}\label{eq:profileu}\\
&&{d\over d\rho}{|b|^2\over 2}={d\over d\rho}{B^2\over 2}=(M_a^2\VV^2-\beta).\label{eq:profileb}
\end{eqnarray}
Equation (\ref{eq:profileb}) can also be rewritten as
$dP/d\rho=\VV^2$, where $P$ denotes the total pressure
${|b|^2\over 2M_a^2}+{\rho\over M_s^2}$.
Equations (\ref{eq:fs-speeds}) and
(\ref{eq:profileu})-(\ref{eq:profileb}) can be solved numerically
but it is advantageous to search for asymptotic solutions in some
limits. First of all, when $\beta=0$, i.e. in the absence of thermal
pressure, it is found that $\VV_-=0$ (the slow wave does not
propagate) and $\VV_+^2={B^2\over M_a^2\rho}$, which is the Alfv\'en
speed based on the total magnetic field intensity. When $b_x=0$,
i.e. for a propagation perpendicular to the ambient magnetic field,
we again have $\VV_-=0$ and now $\VV_+^2={B^2+\beta\rho\over
M_a^2\rho}$. 
In both limits, for the slow waves $P={\rm const}$, while for the
fast waves $B^2 \propto \rho^2$. These relations are in fact more
general as will be seen below.

In the case where $4\beta b_x^2\rho \ll (B^2+\beta\rho)^2$ one
obtains
\begin{eqnarray}
&&\VV_-^2= {b_x^2\over M_s^2
(B^2+\beta\rho)}\label{eq:slowspeed}\\
&&\VV_+^2={B^2\over M_a^2\rho}+{1\over M_s^2} .\label{eq:fastspeed}
\end{eqnarray}
The above assumption only fails
when $\xi={\beta\rho\over b_x^2}$ is of order unity together with
$\eta={|b|^2\over b_x^2}$ small, i.e., when $b_x$ is not too small, for
$\beta\rho$ of order unity and small field distortions. As an
example, the above approximations fail for sound waves propagating
along the magnetic field, for which arbitrary density pertubations
can develop on an unperturbed magnetic field.

Equation (\ref{eq:profileb}) together with the approximation
(\ref{eq:fastspeed}) leads to $B^2 \propto \rho^2$ for the fast
wave. Using  eq. (\ref{eq:slowspeed}) for the slow speed, 
one is led to the following ordinary differential equation
\begin{equation}
{d\eta\over 2d\xi}={1\over 1+\xi+\eta}-1,
\end{equation}
whose implicit solution reads
\begin{equation}
\xi+{\eta\over 2}+\ln |1-\xi-\eta|=C',
\end{equation}
$C'$ being an arbitrary constant.
For large enough $\xi$ and $\eta$, the solution can be approximated
by $\eta=C''-2\xi$. In physical terms it can be rewritten as
$P={\rm const}$. For smaller values of $\xi$ and $\eta$ it can
be shown that $P$ is an increasing function of $\rho$, with
slope $1/2$ when $\eta (\xi=0) =1$, and even larger slope at smaller
values of $\xi$ and $\eta$.

In order to interpret the numerical simulations of Section 4, it is
also useful to investigate under which conditions do the slow or the
fast waves dominate the production of density fluctuations.
%which, among the fast and slow waves, is the
%most effective at producing density fluctuations.
A criterion that can be used  is the relation between the total
velocity vector ${\bf U}=(u,v_y,v_z)$ of the perturbation (in some
way related to the 
fluid displacement) and the density perturbation. From
eqs. (\ref{eq:simple1})-(\ref{eq:fs-speeds}), it
follows that, for small enough perturbations, the density
fluctuation $\Delta \rho$ and the 
velocity fluctuation $\Delta {\bf U_{\pm}}=(\Delta u_{\pm},\Delta
v_{y\pm},\Delta v_{z\pm})$  are related by
\[ 
\Delta u_{\pm} = \VV_{\pm}\frac{ \Delta\rho} {\rho}, 
\]
and
\[
\Delta v_{\pm}=-\frac{b_x b}{\VV_{\pm} M_a^2 |b|^2} (\VV_{\pm}^2
M_a^2 -\beta) \frac{\Delta \rho}{\rho}  .
\]
These relations are not valid for circularly polarized Alfv\'en
waves where $|b|$ is constant.
Using eq. (\ref{eq:fs-speeds}), it is easy to verify that 
$\Delta {\bf U_+} \cdot \Delta {\bf U_-} =0$.
In the linear case, i.e. when $b_y=0$, $b_z=\sin \theta$ and $\rho=1$,
one finds that, in the limit $\beta \rightarrow 0$ 
\[ \left\{ \begin{array}{ll}
%\begin{eqnarray*}
\Delta {\bf U_+} &\propto \sin\theta {\bf \hat x} -\cos \theta
{\bf \hat z} \qquad
\perp {\bf B_0} \\
\Delta {\bf U_-} &\propto \cos\theta {\bf \hat x} +\sin \theta {\bf \hat
  z } \qquad \parallel \ {\bf B_0}
\end{array} \right. 
\]
%\end{eqnarray*}

\noindent while in the limit $\beta \rightarrow \infty$
\[ \left\{ \begin{array}{ll}
%\begin{eqnarray*}
\Delta {\bf U_+} &\propto  {\bf \hat x} \qquad \parallel {\bf k} \\
\Delta {\bf U_-} &\propto  {\bf \hat z} \qquad \perp  {\bf k}
\end{array} \right. 
\] 
%\end{eqnarray*}
where ${\bf k}$ is the wavevector of the linear wave.

Denoting $X_{\pm}=\VV^2_{\pm} M_a^2$ and using the dispersion relation 
$\rho X_{\pm}^2 -(B^2+\beta\rho)X_{\pm}+\beta b_x^2=0$, one
finds 

\begin{equation}
|\Delta U_{\pm}|^2 = \frac{1}{M_a^2 |b|^2}\left ( B^2
  X_{\pm}+\beta\rho X_{\mp} -2\beta b_x^2\right )\left ( \frac{\Delta
    \rho} {\rho} \right )^2  .
\end{equation}
In the limit $\beta \rightarrow 0$ one has $X_+=\frac{B^2}{\rho}$
and $X_-=\frac{\beta b_x^2}{B^2}$ so that
\begin{eqnarray*}
&& |\Delta {\bf U_+}|^2= \frac{B^4}{\rho M_a^2 |b|^2} \left ( \frac{\Delta
    \rho} {\rho} \right )^2 \\
&& |\Delta {\bf U_-}|^2=  \frac{\beta}{M_a^2} \left ( \frac{\Delta
    \rho} {\rho} \right )^2
\end{eqnarray*}
while in the limit $\beta \rightarrow \infty$,
 $X_+=\beta$, $X_-=\frac{b_x^2}{\rho}$ and
\begin{eqnarray*}
&& |\Delta {\bf U_+}|^2= \frac{\beta}{M_a^2} \left ( \frac{\Delta
    \rho} {\rho} \right )^2 \\
&& |\Delta {\bf U_-}|^2=  \frac{\beta^2\rho}{M_a^2 |b|^2} \left ( \frac{\Delta
    \rho} {\rho} \right )^2 .
\end{eqnarray*}
\\From these relations one can conclude that at small 
$\beta$, density fluctuations are mostly created by the slow mode
(since small values of $|\Delta {\bf U_-}|$ are sufficient to
obtain $\frac{\Delta \rho}{\rho} \sim 1$). 
Conversely, the fast mode can more easily generate density
fluctuations at large $\beta$. For intermediate values of $\beta$, 
conclusions may depend on the magnitude of $b_z$. For example, at
parallel propagation and large field distortions, the slow mode
tends to be the most efficient to produce density fluctuations,
except at large density.

\section{Alfv\'en wave pressure} \label{sec:alfven}

In the previous Section we considered waves propagating
on a uniform background. In a turbulent medium, the situation is
usually more complex, as all types of waves are mixed. Another case
that still remains simple enough to be analytically
tractable, consists in studying the properties of MHD waves
propagating on top of a circularly polarized, parallel-propagating
Alfv\'en wave. These Alfv\'en waves are exact solutions of the MHD
equations (Ferraro, 1955) and can be taken of arbitrary amplitude.
They read $b_0=B_0\exp\left[ {-i\sigma (k_0x-\omega_0 t)}\right]$
with $\rho_0=1$, 
$b_{x0}=1$, $u_0=0$ and $v_0=-b_0/M_a$. The polarization of the wave
is determined by the parameter $\sigma$, with $\sigma=+1$ ($\sigma=-1$) for a
right-handed (left-handed) wave. The dispersion relation reads
$\omega_0 =k_0/M_a$. 

Let us now consider perturbations of the form
\begin{eqnarray}
&&b=b_0+b_+e^{-i\sigma \left[ (k+k_0)x-(\omega+\omega_0)t \right]}\nonumber\\
&&+ b_-e^{+i\sigma \left[ (k-k_0)x-(\omega-\omega_0)t \right]}\\
&&\rho=1+\rho'e^{-i\sigma (kx-\omega t)}+c.c.\\
&&u=u'e^{-i\sigma (kx-\omega t)}+c.c.\\
&&v=v_0+v_+e^{-i\sigma \left[ (k+k_0)x-(\omega+\omega_0)t \right]}\nonumber\\
&&+ v_-e^{+i\sigma \left[ (k-k_0)x-(\omega-\omega_0)t \right]}
\end{eqnarray}

The linearized equations read
\begin{eqnarray}
&&\omega \rho'-ku'=0\\
&&\omega u'= {k\over M_s^2}\rho'+{k\over 2 M_a^2} (b_0b_-^*+b_0^*b_+)\\
&&(\omega+\omega_0)v_++{k_0b_0\over M_a}u'-{\omega_0b_0\over M_a}\rho'={-1\over
M_a^2} (k+k_0)b_+\\
&&(\omega-\omega_0)v_--{k_0b_0\over M_a}u^{'*}+{\omega_0b_0\over M_a}\rho^{'*}
= {-1\over M_a^2} (k-k_0)b_-\\
&&(\omega+\omega_0)b_+-(k+k_0)b_0u'=-(k+k_0)v_+\\
&&(\omega-\omega_0)b_--(k-k_0)b_0u^{'*}=-(k-k_0)v_-\ .
\end{eqnarray}
After some algebra, we obtain the following dispersion relation
\begin{eqnarray}
&& \left({k^2\over M_s^2}-\omega^2\right)\left({k_+^2\over M_a^2}-
\omega_+^2\right) \left({k_-^2\over M_a^2}-\omega_-^2\right)+
{k^2B_0^2\over M_a^2} \times \nonumber\\
&&\left(\omega-{k\over M_a}\right)\left(\omega^3+{k\over
M_a}\omega^2 - 3{k_0^2\over M_a^2}\omega+{kk_0^2\over M_a^3}\right)=0.
\end{eqnarray}
Three different limiting cases can be considered and easily
identified after rewriting the dispersion relation as
\begin{eqnarray}
&& \omega^2={k^2\over M_s^2}+\nonumber\\
&& {k^2B_0^2\over M_a^2}
{\left(\omega-{k\over M_a}\right) \left(\omega^3+{k\over
M_a}\omega^2 -3{k_0^2\over M_a^2}\omega+{kk_0^2\over M_a^3}\right)
\over \left({k_+^2\over M_a^2}-\omega_+^2\right)
\left({k_-^2\over M_a^2}-\omega_-^2\right)}.
\end{eqnarray}

In the case $M_a\rightarrow \infty$, i.e. in the case of a weak
background magnetic field, magnetic tension is negligible compared
to field stretching and the dispersion relation approximates to 
$\omega^2\approx{k^2\over M_s^2}+{k^2B_0^2\over M_a^2}$.
If one assumes a polytropic dependence of the magnetic pressure on
the density in the form $|b|^2/2 \propto \rho^{\gamma}$ and
linearizes eqs. (\ref{eq:mhd1d1})-(\ref{eq:mhd1d2}), a direct
comparison of the resulting dispersion relation with the above
approximation leads to $\gamma=2$. This is the case, mentioned by 
\cite{MZ95}, in which the perturbation is very rapid
compared to the speed of the Alfv\'en wave. This behavior of the
magnetic pressure is easily analyzed directly from the orginal MHD
equations. Indeed, in the limit where $M_a$ is very
large, the term on the right-hand-side of eq. (\ref{eq:mhd1d4}) can
be dropped and this equation leads to 
\begin{equation}
{\partial |b|^2\over\partial t}+ {\partial\over \partial
x}(u|b|^2) +|b|^2 {\partial\over \partial x} u=0.
\end{equation}
Together with eq. (\ref{eq:mhd1d1}), one obtains
\begin{equation}
({\partial \over\partial t}+u{\partial\over \partial x}) (\ln
|b|^2-2\ln \rho)=0,
\end{equation}
confirming the result $\gamma=2$ in this case.

In the case $M_a\rightarrow 0$, i.e. in the case of a strong
background magnetic field, magnetic tension is dominant and the
Alfv\'en wave is very rapid. When $\omega\approx 0$ and $k\ll k_0$,
i.e. when the perturbation is very slow, or ``quasi-static'', the
dispersion relation approximates to $\omega^2\approx{k^2\over
M_s^2}+{k^2B_0^2\over 4 M_a^2}$, giving $\gamma_e={1\over 2}$,
also derived by \cite{MZ95} using a WKB approximation. 

Finally, for $M_a$ finite but $k\approx 0$ and $\omega\ll \omega_0$,
i.e. for quasi-uniform perturbations,  the
dispersion relation approximates to $\omega^2\approx{k^2\over
M_s^2}+{3 k^2B_0^2\over 4 M_a^2}$, giving $\gamma_e={3\over 2}$,
recovering the prediction of \cite{MZ95} for this case.

Whereas the predictions based on the WKB approach are recovered in
this purely one-dimensional linear analysis, it should be noted that
they are probably not valid in a fully
three-dimensional situation. For example, 
\cite{MZ95} argued that the Alfv\'en wave pressure is
isotropic.  A linear stability analysis analogous to the previous
one but performed in the long-wave limit and for perturbations
propagating exactly perpendicular to the background Alfv\'en wave
shows that the Alfv\'en wave exerts a pressure 
different from the 1D case, thus invalidating the isotropy result. 
This pressure can even be negative
and one recovers the polytropic index $\gamma_e =3/2$ only in the
limit of large amplitude background wave and for $\beta=0$
(Passot \& Gazol, in preparation).

\section{Numerical simulations} \label{sec:num_sim}

In order to test the range of validity of the  analytical
predictions, we have performed numerical simulations of
eqs. (\ref{eq:mhd1d1})-(\ref{eq:mhd1d4}) in a periodic domain  using
a pseudo-spectral method based on Fourier expansions.  In order to
numerically handle the formation of strong shocks, dissipative terms
of the  usual form, namely
${\mu_l\over \rho} {\partial^2 u\over \partial x^2}$ and 
${\mu_t\over \rho} {\partial^2 v\over \partial x^2}$, are added to
the right-hand-side of eqs.  (\ref{eq:mhd1d2}) and (\ref{eq:mhd1d3}),
respectively, and  a magnetic diffusion 
$\eta {\partial^2 b\over \partial x^2}$ to the
right-hand-side of eq.  (\ref{eq:mhd1d4}). 
In addition, it was found necessary to add  a mass diffusion in
the form $\mu_r {\partial^2 \rho \over \partial x^2}$ to the
right-hand-side of eq. (\ref{eq:mhd1d1}). This term
preserves mass conservation and allows handling of strong shocks. If
kept small enough, it does not significantly modify the
dynamics and in particular it does not alter the statistical
conclusions we shall present. Except for one set of simulations
discussed below, the
forcing (actually an acceleration) acting on these equations is
applied on Fourier modes 1-19, peaked at wavenumber 8 with
amplitudes proportional to $k^4 \exp(-k^2/32)$ and
Gaussian-distributed random phases. A forcing is also 
applied on mode 0 in order to ensure momentum conservation.
The random phases are changed every 0.003 time units. A state of
constant density and zero velocity and magnetic field fluctuations
is taken as initial conditions. A resolution of 4096 grid points is
used and typical values for the diffusion coefficients are
$\mu_l=\mu_t=\eta=0.003$ and $\mu_r=6.\times 10^{-4}$.
In the simulations, three main parameters
are varied, namely the angle $\theta$ and the Mach numbers $M_s$ and $M_a$,
but only $\theta$ and $M_a$ are actually important as we only consider
the high-$M_{\rm s}$ limit. 
For each simulation, we compute r.m.s. values 
of the sonic Mach number $\tilde M_s= M_s \left <
\bar U^2 \right >^{1/2}$, where the brackets (resp. bars) denote
time (resp. spatial) averaging. Similarly, the r.m.s. Alfv\'enic Mach
number and the effective beta of the flow are defined as  $\tilde M_a= M_a
\left < 
\overline { \frac{\rho U^2}{B^2} }\right >^{1/2}$ and $\tilde \beta
= \frac{M_a^2}{M_s^2} \left < \overline {\frac{\rho}{B^2}} 
\right > $. Finally, the density fluctuations
and field distortions are defined as ${\tilde \frac{\delta\rho}{\rho}}=\left <
\overline {(\rho -1)^2}\right >^{1/2}$ and ${\tilde \frac{\delta
B}{B}} = \left <\overline {(b_y^2 +(b_z-\sin\theta)^2)^2}\right
>^{1/2}$ respectively. 
\begin{figure}[htb]
\centerline{\hbox{
\psfig{figure=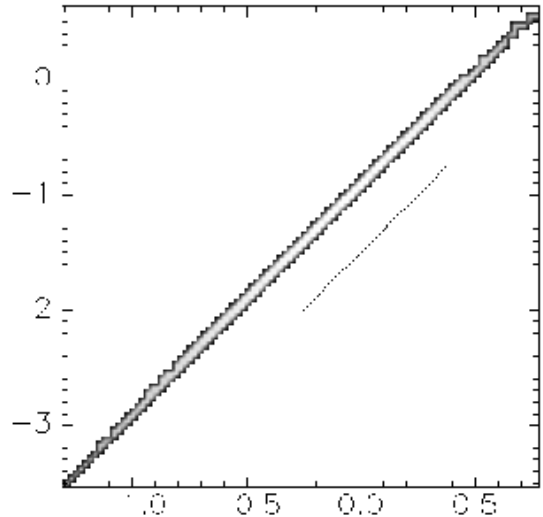,bbllx=4cm,bblly=8cm,bburx=16.cm,bbury=20.cm,height=8.cm,width=8.cm,clip=}}}
\caption{Log-log scatter plot of the magnetic pressure versus
density for a simulation with $\theta=\pi/2$, $\tilde M_s=4.12$,
$\tilde M_a=5.16$, $\tilde \beta = 1.52$, ${\tilde
\frac{\delta\rho}{\rho}}=0.623$, ${\tilde \frac{\delta
B}{B}} =0.618$ and $\tilde\sigma=0.054$. The line segment has a
slope equal to 2.} 
\label{fig1}
\end{figure}

The simplest case corresponds to perpendicular propagation
($\theta=\pi/2$) where only fast waves can propagate. In that case,
we expect magnetic pressure to be  very well correlated with
density, with a dependance of the form $|b|^2 \propto \rho^2$,
whatever the value of $M_a$.
This prediction is confirmed by the simulations as examplified in
Fig. 1, which shows the logarithm of the 
two-dimensional histogram $h\left[\ln(\rho),\ln(|b|^2/2M_a^2)\right]$
 of the magnetic pressure versus density
for a run with $\tilde M_a=5.16$, adding points from 
temporal snapshots taken every $0.1$ time units from $t=100$ to
$t=10^4$. More precisely we display in grey scale (with black
(resp. white) denoting the smallest (resp. largest) value) the
logarithm of the number of points (in the space-time
sample) in a given interval of $|b|^2/2M_a^2$ and
$\rho$, plotting only the points where the histogram is larger
than $90\%$ of the maximum at fixed $\rho$. We also
calculate an average 
histogram dispersion in the following way. For each density $\rho$,
we define the mean of the magnetic pressure logarithm as 
\begin{equation}
\bar p_m (\rho) = \int_0^\infty x h\left[\ln(\rho),x\right] dx /H
\end{equation}
and the dispersion as
\begin{equation}
\bar \sigma (\rho) = \left( \int_0^\infty \left[x-\bar p_m(\rho)\right]^2
h\left[\ln(\rho),x\right] dx/H \right ) ^{1/2} 
\end{equation}
where $H=\int_0^\infty h(\rho,x) dx$.
An average dispersion $\tilde \sigma$ is evaluated by taking the
mean of $\bar \sigma (\rho)$ for $\ln(\rho)$ varying by $10\%$ about its
central value.

\begin{figure}[htb]
\centerline{\hbox{
\psfig{figure=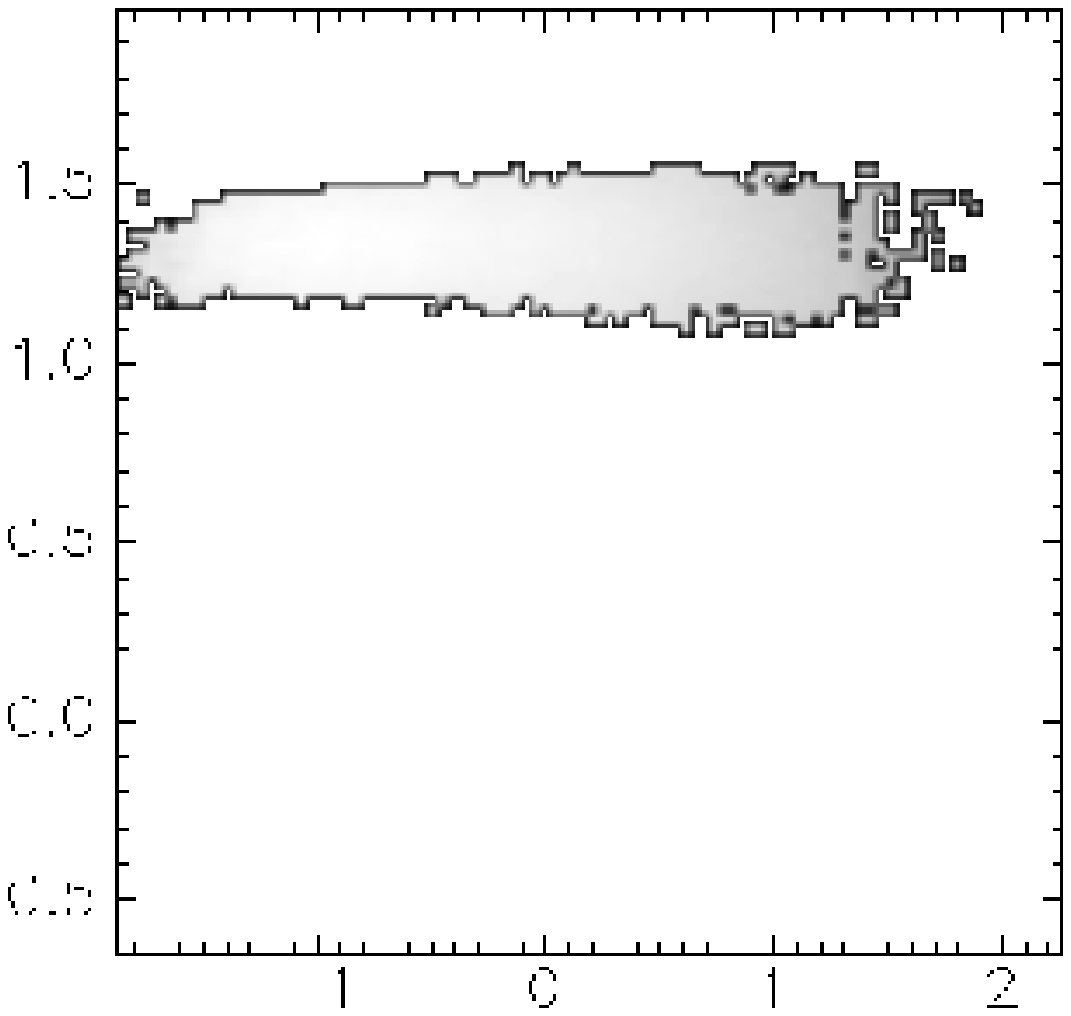,bbllx=4cm,bblly=8cm,bburx=16.cm,bbury=20.cm,height=4.cm,width=4.cm,clip=},
\psfig{figure=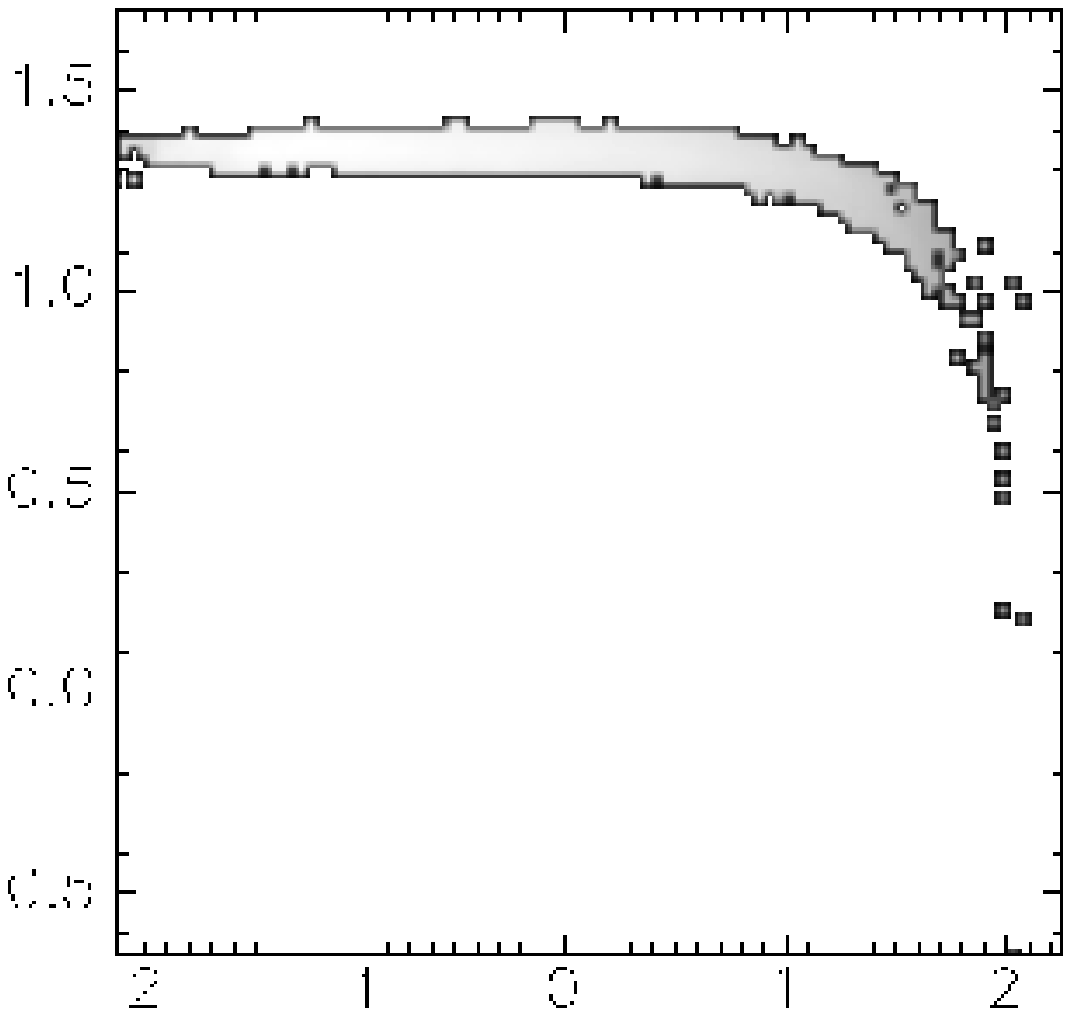,bbllx=4cm,bblly=8cm,bburx=16.cm,bbury=20.cm,height=4.cm,width=4.cm,clip=}}}
\caption{Log-log scatter plot of the magnetic pressure versus
density for two simulations with $\cos\theta=0.1$. The parameters of
the left panel, corresponding to a case with forcing on all three
velocity components  are $\tilde M_s=6.63$,
$\tilde M_a=0.478$, $\tilde \beta = 0.00685$, ${\tilde
\frac{\delta\rho}{\rho}}=2.69$, ${\tilde \frac{\delta
B}{B}} =0.323$ and $\tilde\sigma=0.18$. For the right panel, where the
forcing is applied 
on the $v$ velocity components only, $\tilde M_s=7.27$,
$\tilde M_a=0.527$, $\tilde \beta = 0.00728$, ${\tilde
\frac{\delta\rho}{\rho}}=3.55$, ${\tilde \frac{\delta
B}{B}} =0.324$ and $\tilde\sigma=0.059$. }
\label{fig2}
\end{figure}

\begin{figure}[htb]
\centerline{\vbox{
\psfig{figure=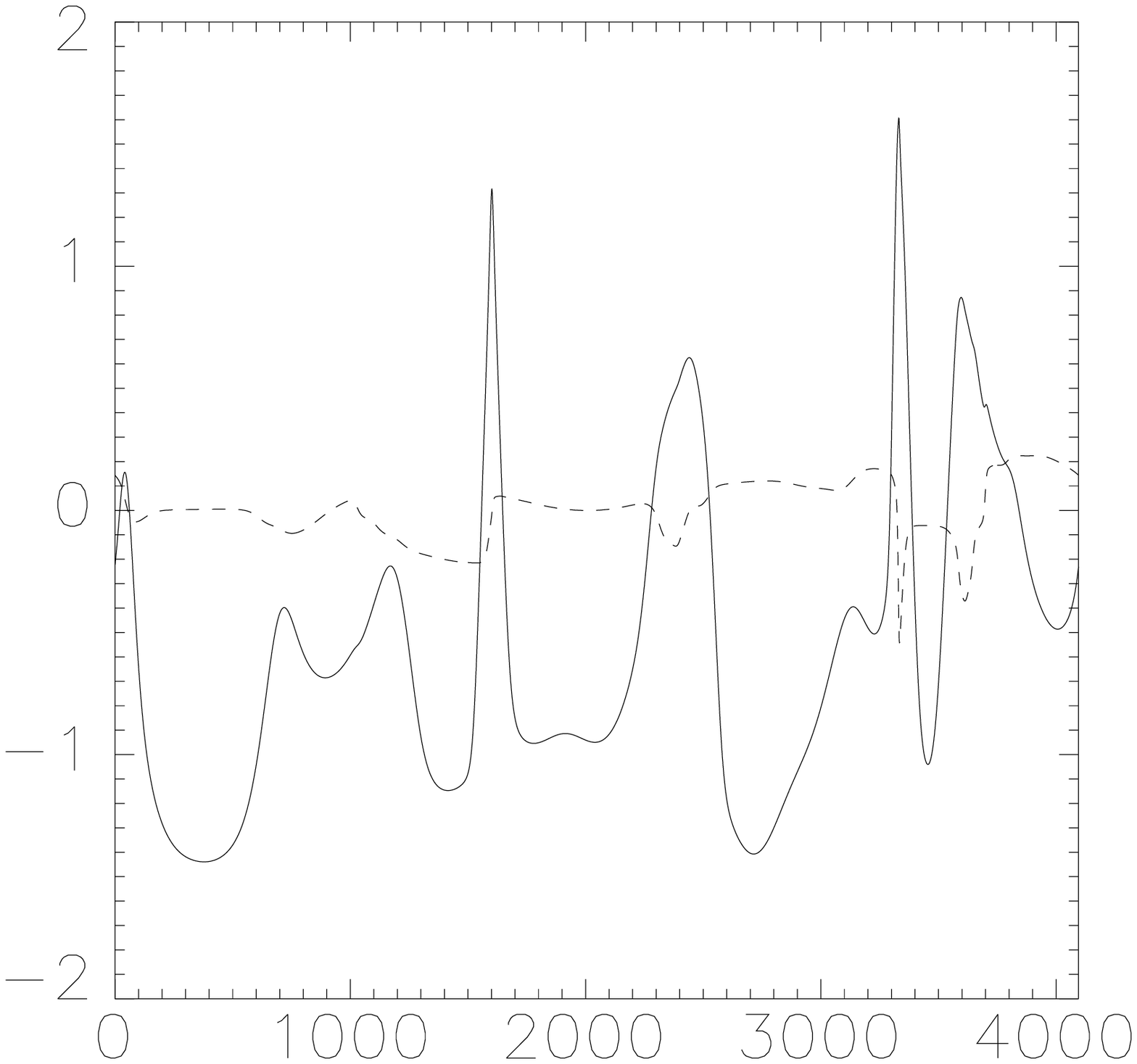,bbllx=1cm,bblly=3cm,bburx=19.cm,bbury=23cm,height=7.cm,width=7.cm,clip=} 
\psfig{figure=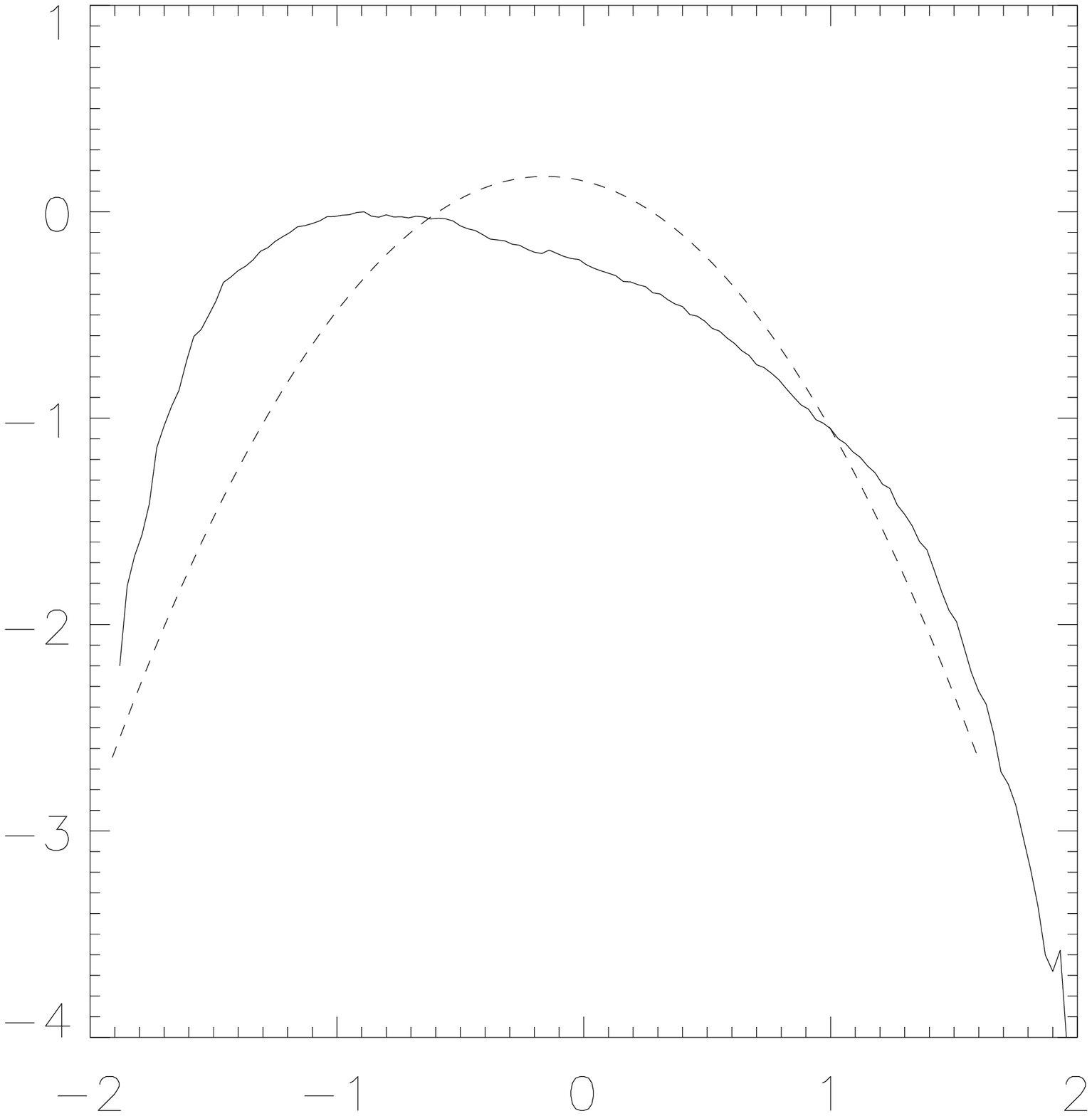,bbllx=.5cm,bblly=4cm,bburx=21.cm,bbury=24.5cm,height=7.cm,width=7.cm,clip=}}}
\caption{{\it Top}: Snapshot of $\log_{10}\rho$ (solid) and $\log_{10}|b|^2$
(dotted) for the run of Fig. 2a. {\it Bottom}: Density PDF for the run of Fig. 2a in log-log
coordinates. The dashed line shows a fit of the curve with
a parabola corresponding, in these coordinates, to a lognormal distribution.}
\label{fig3}
\end{figure}

As the angle $\theta$ is decreased from $\pi/2$, the behavior is
very sensitive to $M_a$. When $M_a \ll 1$ the field lines are still
almost unperturbed but the density fluctuations are very different
from the case $\theta=\pi/2$ because for small $\beta$  the latter are
mostly created by slow waves. In that case again we expect the magnetic
pressure to exhibit little scatter, as the
field is only slightly perturbed, and to be roughly anti-correlated
with density, as the total pressure remains roughly constant.
%a tight
%correlation between magnetic pressure and density with the total
%pressure scaling roughly independently of the density. 
This is indeed the case, as seen in Fig. 2a. where the distribution of
$|b|^2/2M_a^2$ and $\rho$  has a small scatter. 
In addition, a clear anticorrelation is observed
at high density when no forcing is applied on the $x$-component of
the velocity (Fig. 2b). In the case of purely
perpendicular propagation the density structures are those
typically observed in neutral turbulence with a polytropic index
$\gamma =2$, i.e. rather flat-topped structures (plateaux) separated by
shocks that keep interacting, with a preeminence of low density regions
(\cite{PVS98}).
In contrast, for nearly perpendicular propagation and $M_a \ll 1$,
the density structures are composed of large peaks that oscillate
while avoiding collision (Fig. 3a). The probability density function
(PDF) of the density shown in Fig. 3b  shows an extended tail at
large density with a significant excess at small density. Since the
field strength is roughly constant and independent of $\rho$, the
magnetic pressure term in eq. (\ref{eq:mhd1d2}) acts like an
random acceleration whose strength increases as $\rho$ decreases since
$|b|^2$ is roughly constant at small density (see Fig. 2a).
Whereas  the density field of an isothermal gas stirred by a random acceleration
has a lognormal distribution, when the acceleration depends on the
density, the PDF ceases to be lognormal (\cite{PVS98}). The lift up of
the left tail seen in Fig. 3b is thus expected in the present case.

\begin{figure}[htb]
\centerline{\hbox{
\psfig{figure=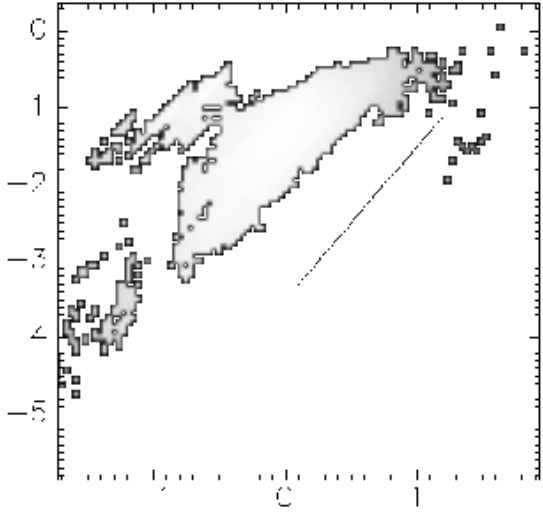,bbllx=4cm,bblly=8cm,bburx=16.cm,bbury=20.cm,height=8.cm,width=8.cm,clip=}}}
\caption{Log-log scatter plot of the magnetic pressure versus
density for a simulation with forcing on the three velocity
components, $\cos\theta=0.1$, $\tilde M_s=5.73$, 
$\tilde M_a=7.29$, $\tilde \beta = 1.53$, ${\tilde
\frac{\delta\rho}{\rho}}=1.32$, ${\tilde \frac{\delta
B}{B}} =2.17$ and $\tilde\sigma=0.61$. The line segment has a slope
equal to 2.} 
\label{fig4}
\end{figure}
\begin{figure}[htb]
\centerline{\vbox{
\psfig{figure=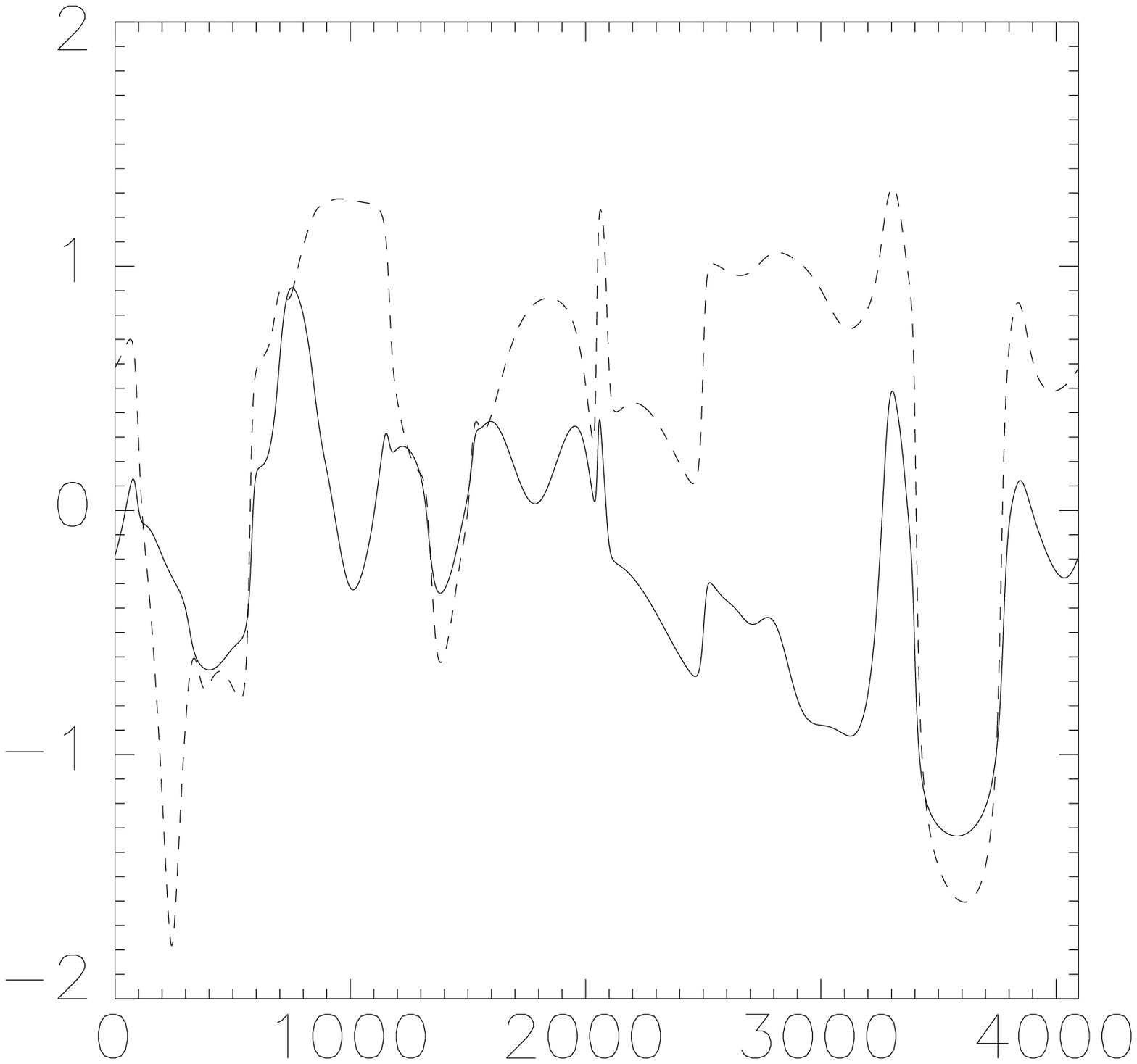,bbllx=1cm,bblly=3cm,bburx=19.cm,bbury=23cm,height=7.cm,width=7.cm,clip=}
\psfig{figure=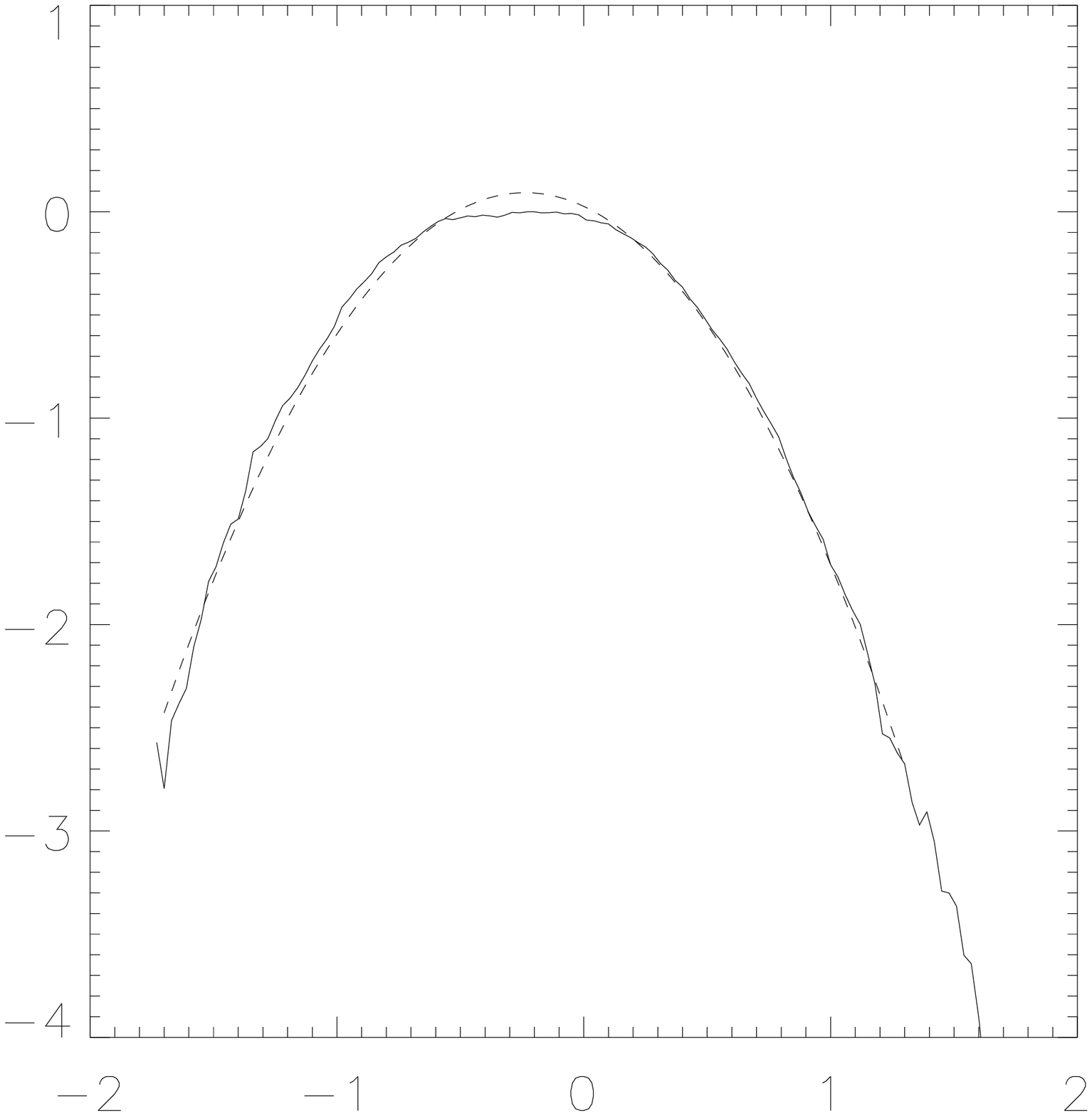,bbllx=.5cm,bblly=4cm,bburx=21.cm,bbury=24.5cm,height=7.cm,width=7.cm,clip=}}}
\caption{{\it Top}: Snapshot of $\log_{10}\rho$ (solid) and $\log_{10}|b|^2$
(dotted) for the run of Fig. 4. {\it Bottom}: Density PDF for the run of Fig. 4 in log-log
coordinates. The dashed shows a lognormal fit.}
\label{fig5}
\end{figure}

When the angle $\theta$ is still slightly smaller than $\pi/2$ but
with $M_a \gg 1$, the magnetic field lines undergo large
fluctuations and both slow and fast waves contribute to the
production of density fluctuations. As seen in Fig. 4, and Fig. 5a,  magnetic
pressure and density are now positively correlated
(magnetic pressure roughly follows a polytropic law with
$\gamma=2$, corresponding to the fast mode), but the dispersion is
larger than in the case of a 
small Alfv\'enic Mach number, indicative of the additional contribution of
the slow mode. 
As a result, the stirring due to magnetic pressure is statistically
non-correlated with the density and  the density PDF is expected to
be much closer to a lognormal as seen in Fig. 5.

\begin{figure}[htb]
\centerline{\hbox{
\psfig{figure=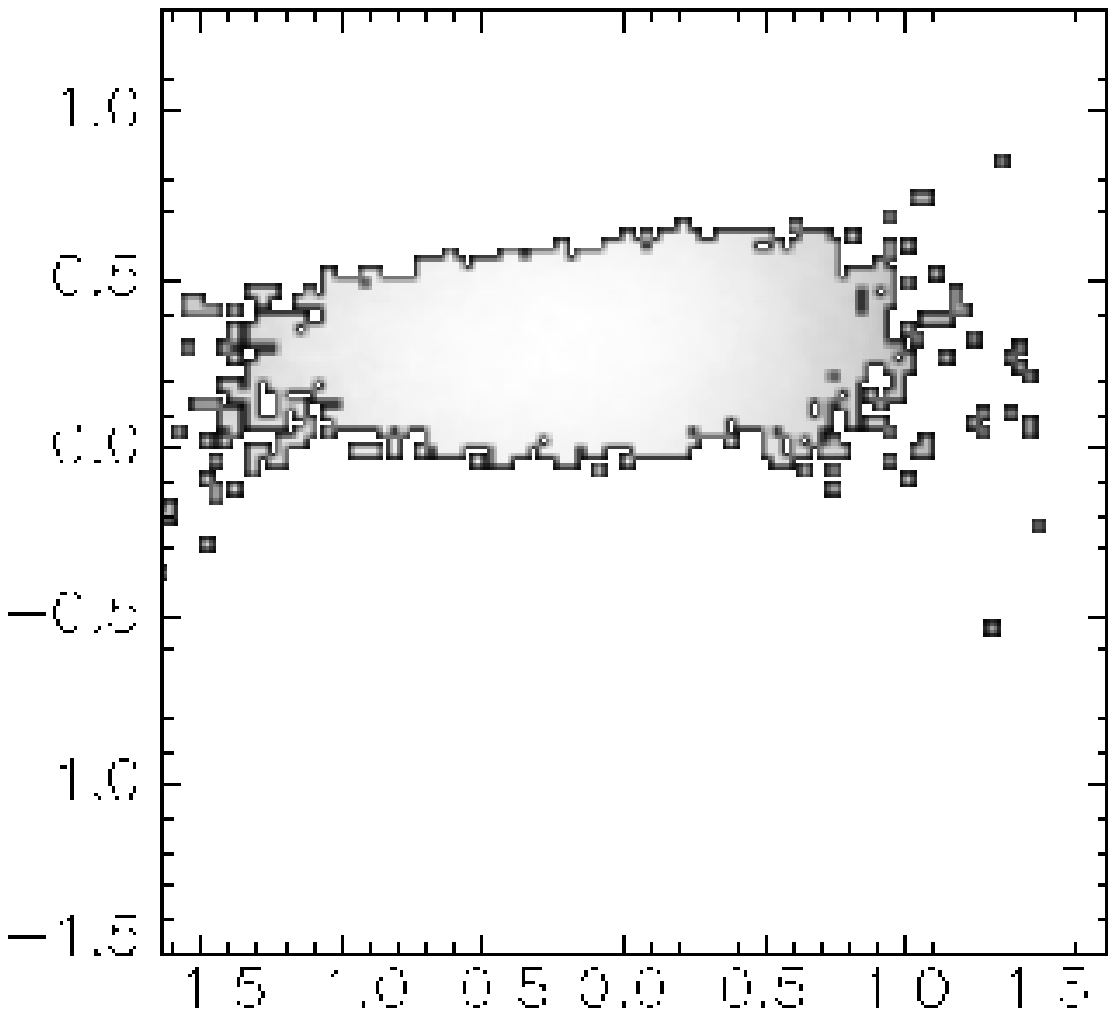,bbllx=4cm,bblly=8cm,bburx=16.cm,bbury=20.cm,height=4.cm,width=4.cm,clip=},
\psfig{figure=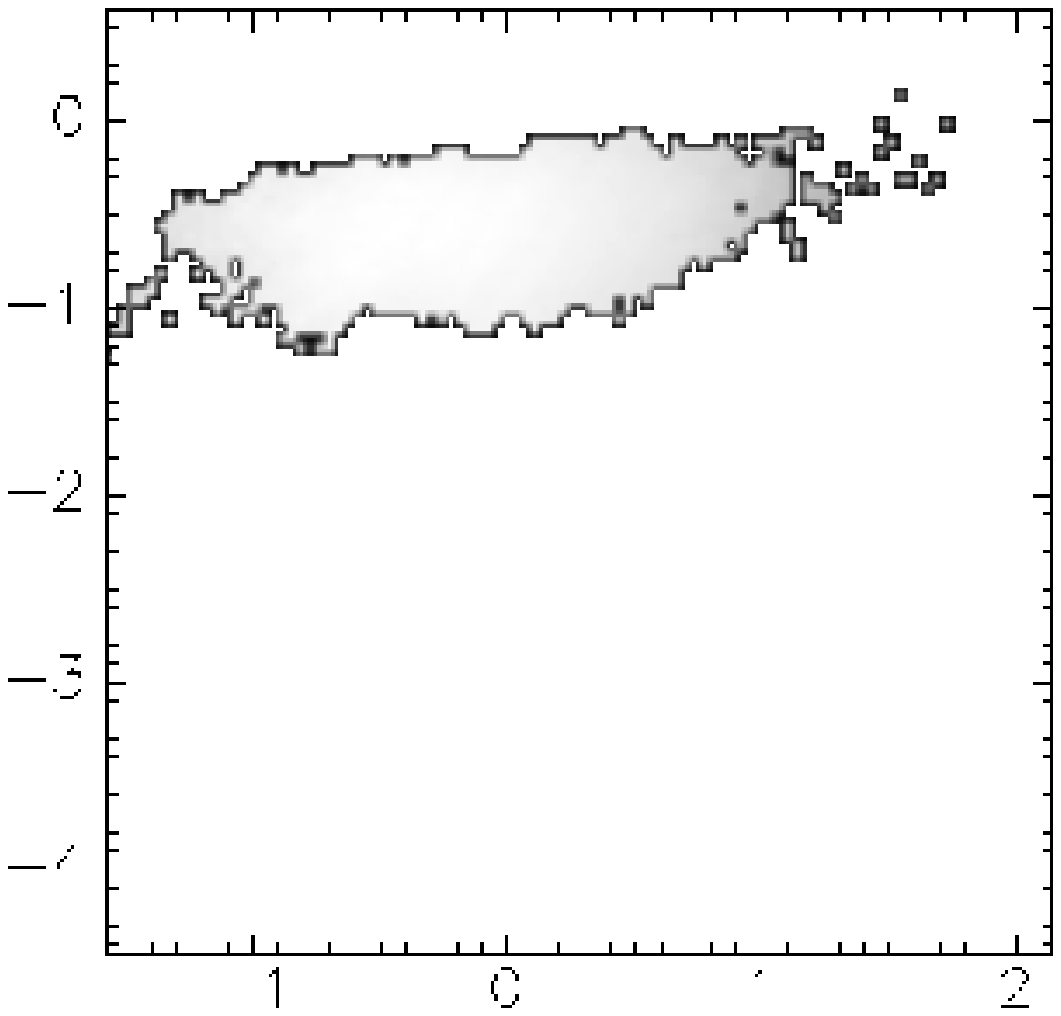,bbllx=4cm,bblly=8cm,bburx=16.cm,bbury=20.cm,height=4.cm,width=4.cm,clip=}}}
\caption{Log-log scatter plot of the magnetic pressure versus
density for two simulations with forcing on the three velocity
components and $\cos\theta=0.7$. The parameters of
the left panel  are $\tilde M_s=2.90$,
$\tilde M_a=0.521$, $\tilde \beta = 0.00375$, ${\tilde
\frac{\delta\rho}{\rho}}=1.20$, ${\tilde \frac{\delta
B}{B}} =0.473$ and $\tilde\sigma=0.32$. For the right panel  $\tilde M_s=6.59$,
$\tilde M_a=2.48$, $\tilde \beta = 0.141$, ${\tilde
\frac{\delta\rho}{\rho}}=1.43$, ${\tilde \frac{\delta
B}{B}} =9.62$ and $\tilde\sigma=0.49$. }
\label{fig6}
\end{figure}

When the angle $\theta$ is intermediate between parallel and almost
perpendicular propagation, the distinction between the
small and the large Alfv\'enic Mach number cases is not as clear. 
As shown in Figs. 6a-b for runs with an  angle $\theta \approx \pi/4$
and for $\tilde M_a= 0.521$ and $\tilde M_a =2.48$ respectively,
there is a larger 
scatter of the points for small to intermediate values of the
density at large Alfv\'enic Mach number. In this case 
a positive correlation between magnetic pressure and
density is noticeable at large density (Fig. 6b).
\begin{figure}[htb]
\centerline{\hbox{
\psfig{figure=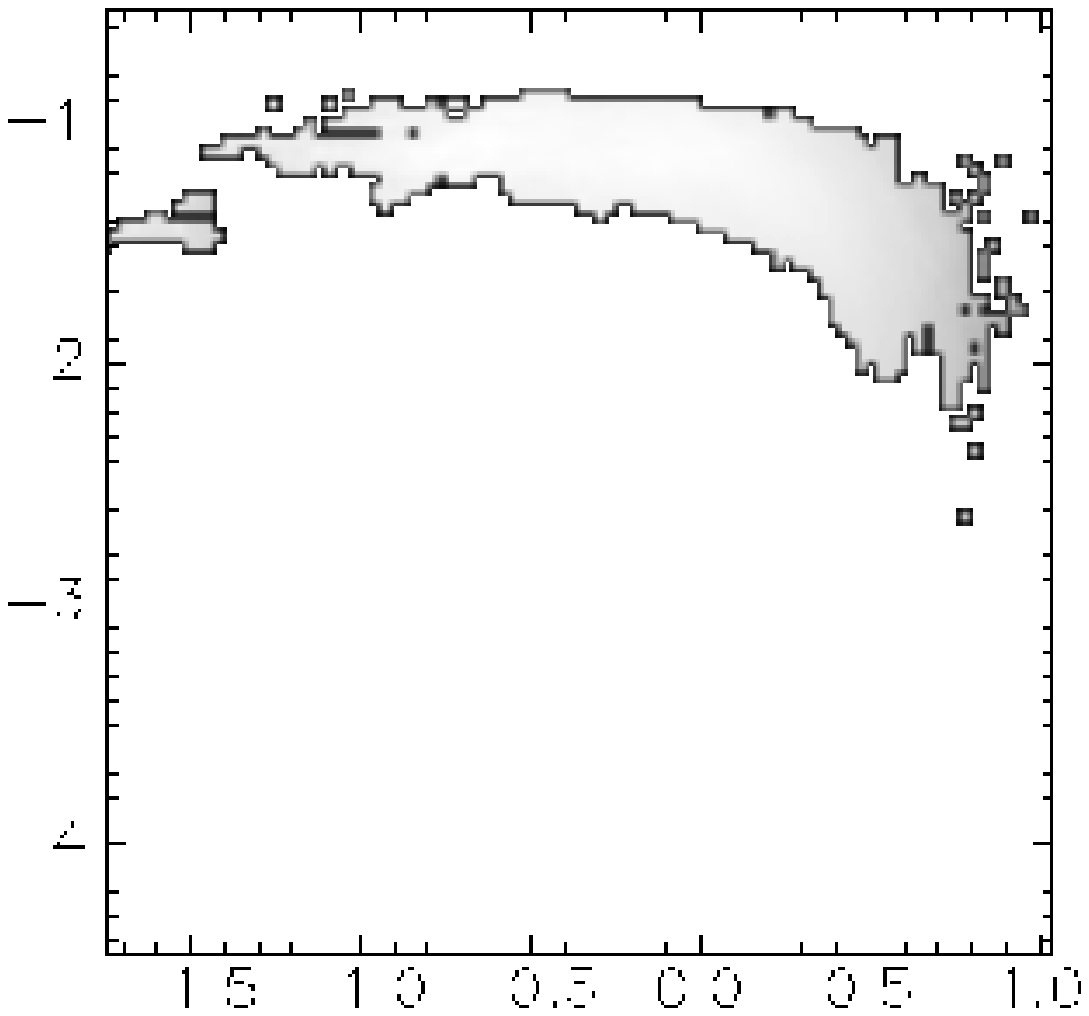,bbllx=4cm,bblly=8cm,bburx=16.cm,bbury=20.cm,height=4.cm,width=4.cm,clip=},
\psfig{figure=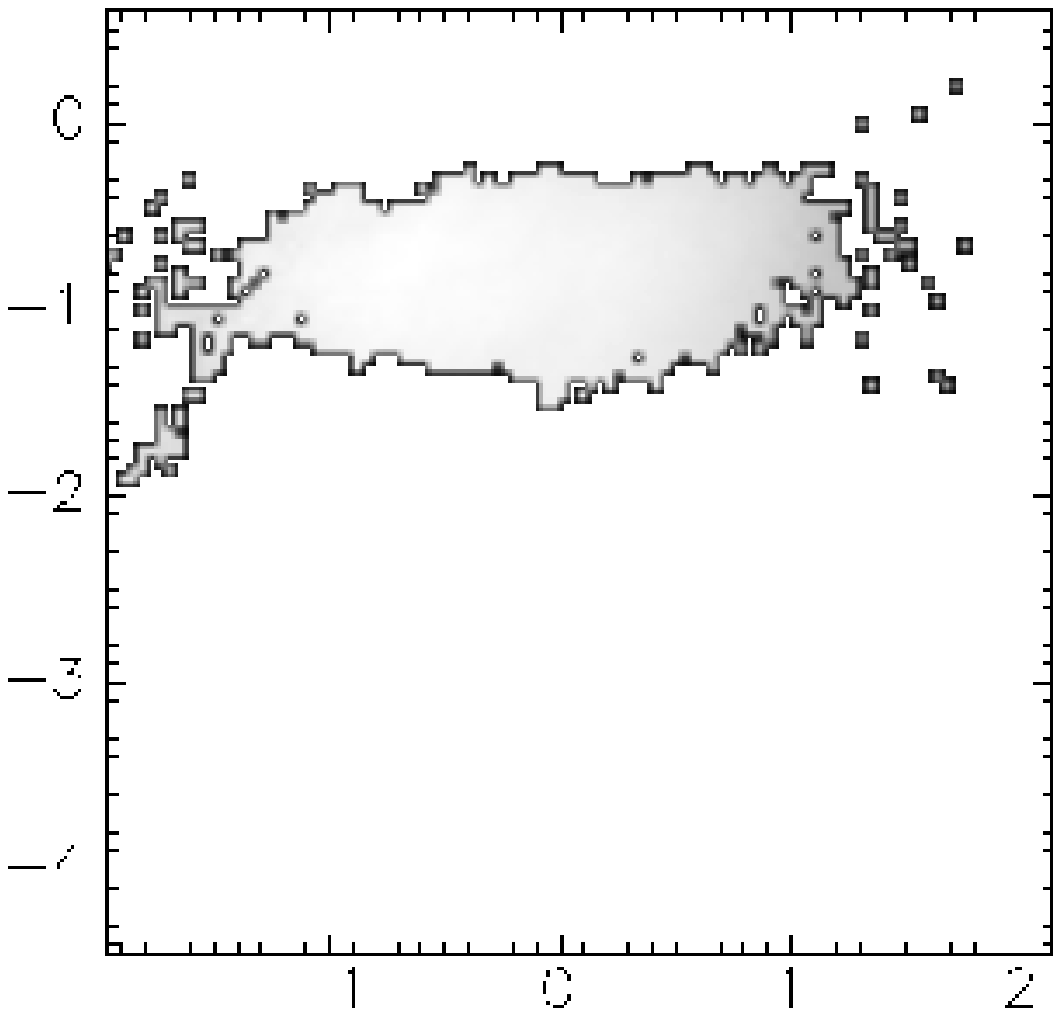,bbllx=4cm,bblly=8cm,bburx=16.cm,bbury=20.cm,height=4.cm,width=4.cm,clip=}}}
\caption{Log-log scatter plot of the magnetic pressure versus
density for two simulations with $\theta=0$. The parameters of
the left panel, where the forcing only affects the perpendicular
velocity components,  are $\tilde M_s=2.72$,
$\tilde M_a=1.76$, $\tilde \beta = 0.440$, ${\tilde
\frac{\delta\rho}{\rho}}=1.04$, ${\tilde \frac{\delta
B}{B}} =4.73$ and $\tilde\sigma=0.21$. For the right panel, where
the forcing is applied on 
the three velocity components, $\tilde M_s=5.10$,
$\tilde M_a=2.36$, $\tilde \beta = 0.218$, ${\tilde
\frac{\delta\rho}{\rho}}=1.73$, ${\tilde \frac{\delta
B}{B}} =7.91$ and $\tilde\sigma=0.57$. }
\label{fig7}
\end{figure}

For parallel propagation with perpendicular forcing, the behavior is
again strongly dependent on the Alfv\'enic Mach number.
When $M_a \gg 1$, the magnetic field is strongly distorted. The
forcing excites slow modes which develop into shocks,
inside which large density clumps form. These clumps, {\it located at
local minima of the magnetic field intensity}, are separated by regions of large
magnetic pressure (Fig. 8a). As a result, they can
hardly approach each other but rather oscillate about their mean
position, at a frequency close to that of the fast wave of the
local total magnetic field. The signature of the slow wave dominance
is the rather small scatter and the anti-correlation between
magnetic pressure and density in compressed regions, as seen in
Fig. 7a corresponding to a case with $\tilde M_a=1.76$. 
In a way similar to the case of small $M_a$ and almost perpendicular
propagation, the density PDF shown in Fig. 8, displays an excess at
small density. 
\begin{figure}[htb]
\centerline{\vbox{
\psfig{figure=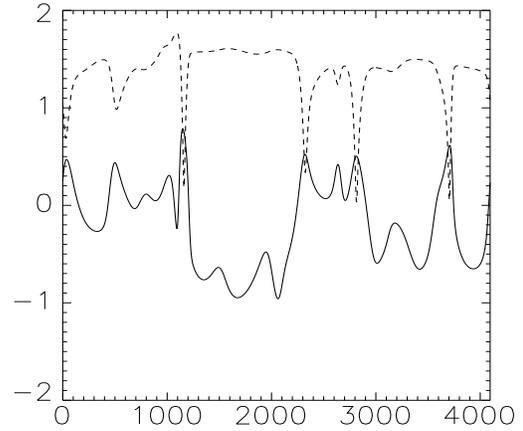,bbllx=1cm,bblly=3cm,bburx=19.cm,bbury=23cm,height=7.cm,width=7.cm,clip=}
\psfig{figure=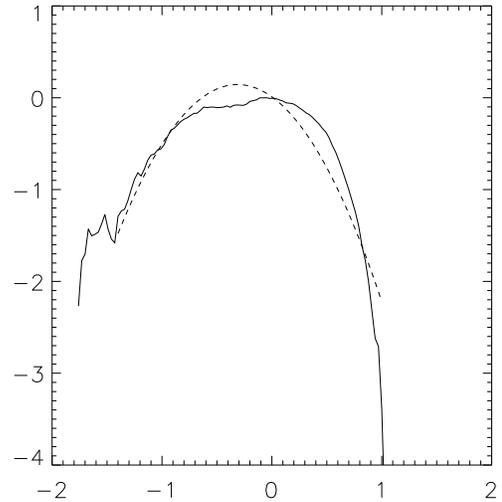,bbllx=.5cm,bblly=4cm,bburx=21.cm,bbury=24.5cm,height=7.cm,width=7.cm,clip=}}}
\caption{{\it Top}: Snapshot of $\log_{10}\rho$ (solid) and $\log_{10}|b|^2$
(dotted) for the run of Fig. 7a. {\it Bottom}: Density PDF for the run of Fig. 7a in log-log
coordinates. The dashed shows a lognormal fit.}
\label{fig8}
\end{figure}

At small values of $M_a$, the magnetic field undergoes mild
variations. The perpendicular forcing preferentially excites fast
waves but the slow waves that form by nonlinear interactions are
the most effective at producing density fluctuations since, the
field being almost straight, only thermal pressure acts against
compression at dominant order (see below for the effect of Alfv\'en
wave pressure). In contrast with the case of large $M_a$, a larger
scatter is observed on the magnetic pressure vs. density diagram,  
as seen on Fig. 9. As seen on Fig. 10a, the correlation between
density and magnetic field intensity is lost and the density PDF is closer to a
lognormal (Fig. 10b).
\begin{figure}[htb]
\centerline{\hbox{
\psfig{figure=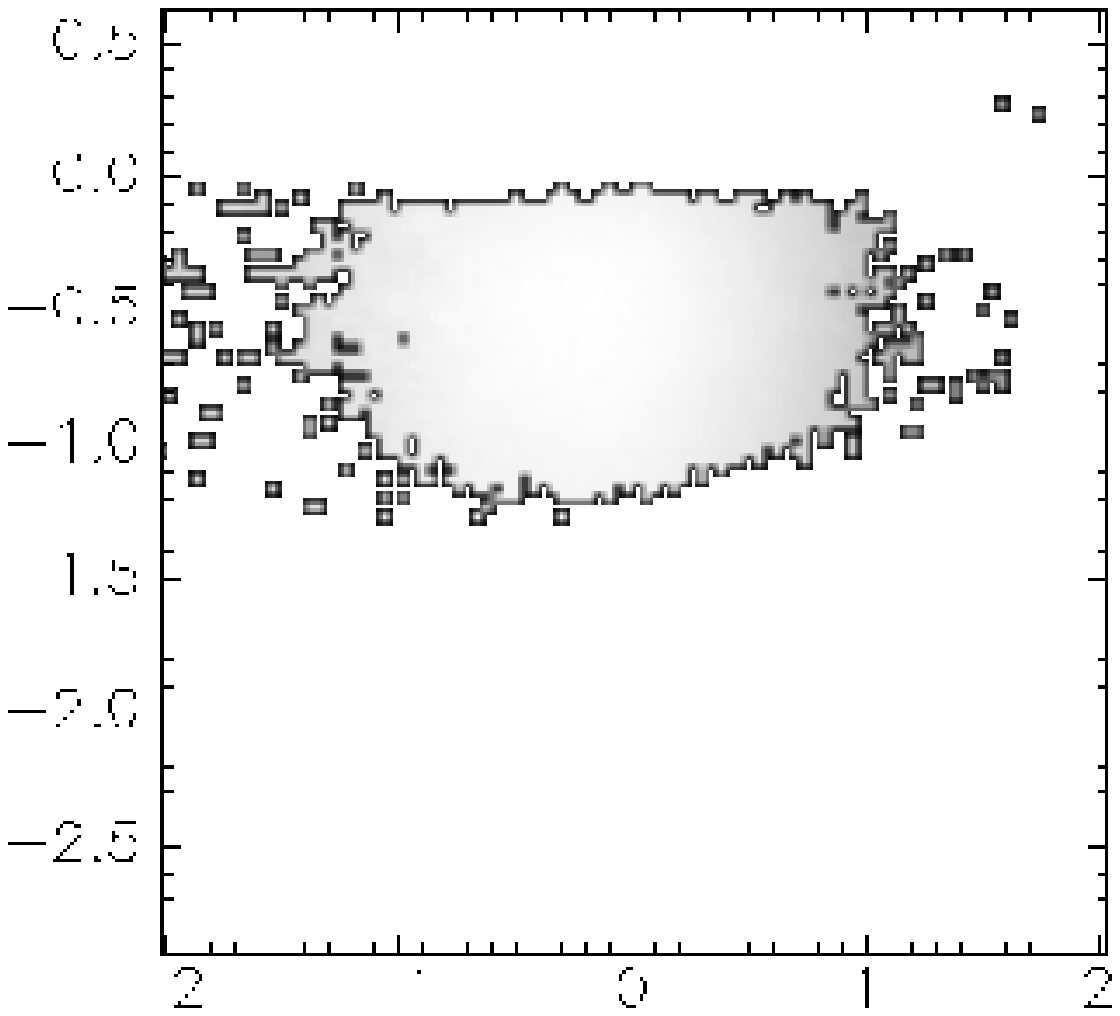,bbllx=4cm,bblly=8cm,bburx=16.cm,bbury=20.cm,height=8.cm,width=8.cm,clip=}}}
\caption{Log-log scatter plot of the magnetic pressure versus
density for a simulation with $\theta=0$ and forcing on the perpendicular
velocity components. The parameters are $\tilde M_s=5.34$,
$\tilde M_a=0.172$, $\tilde \beta = 0.00108$, ${\tilde
\frac{\delta\rho}{\rho}}=1.37$, ${\tilde \frac{\delta
B}{B}} =0.168$ and $\tilde\sigma=0.56$.}
\label{fig4}
\end{figure}
\begin{figure}[htb]
\centerline{\vbox{
\psfig{figure=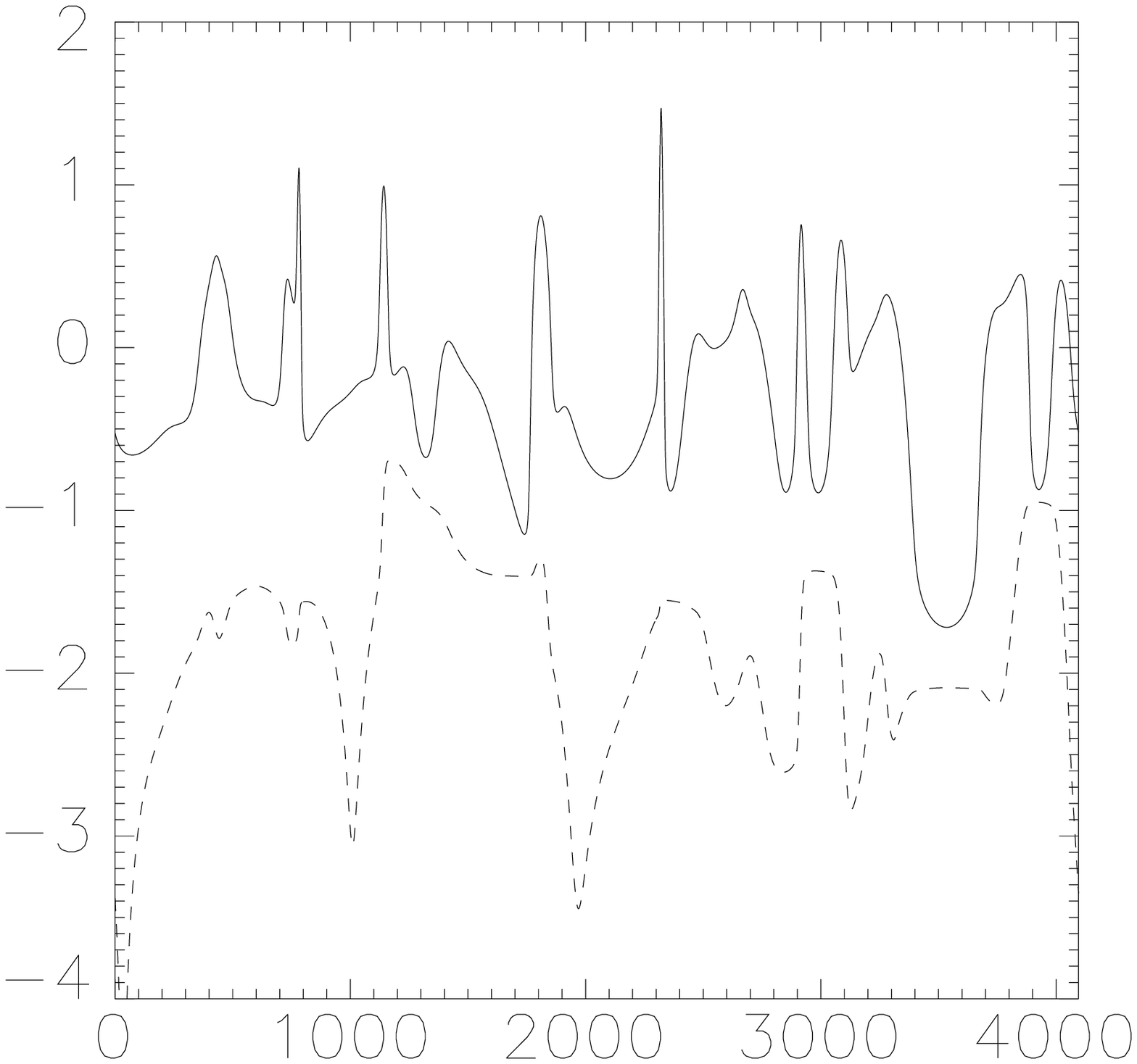,bbllx=1cm,bblly=3cm,bburx=19.cm,bbury=23cm,height=7.cm,width=7.cm,clip=} 
\psfig{figure=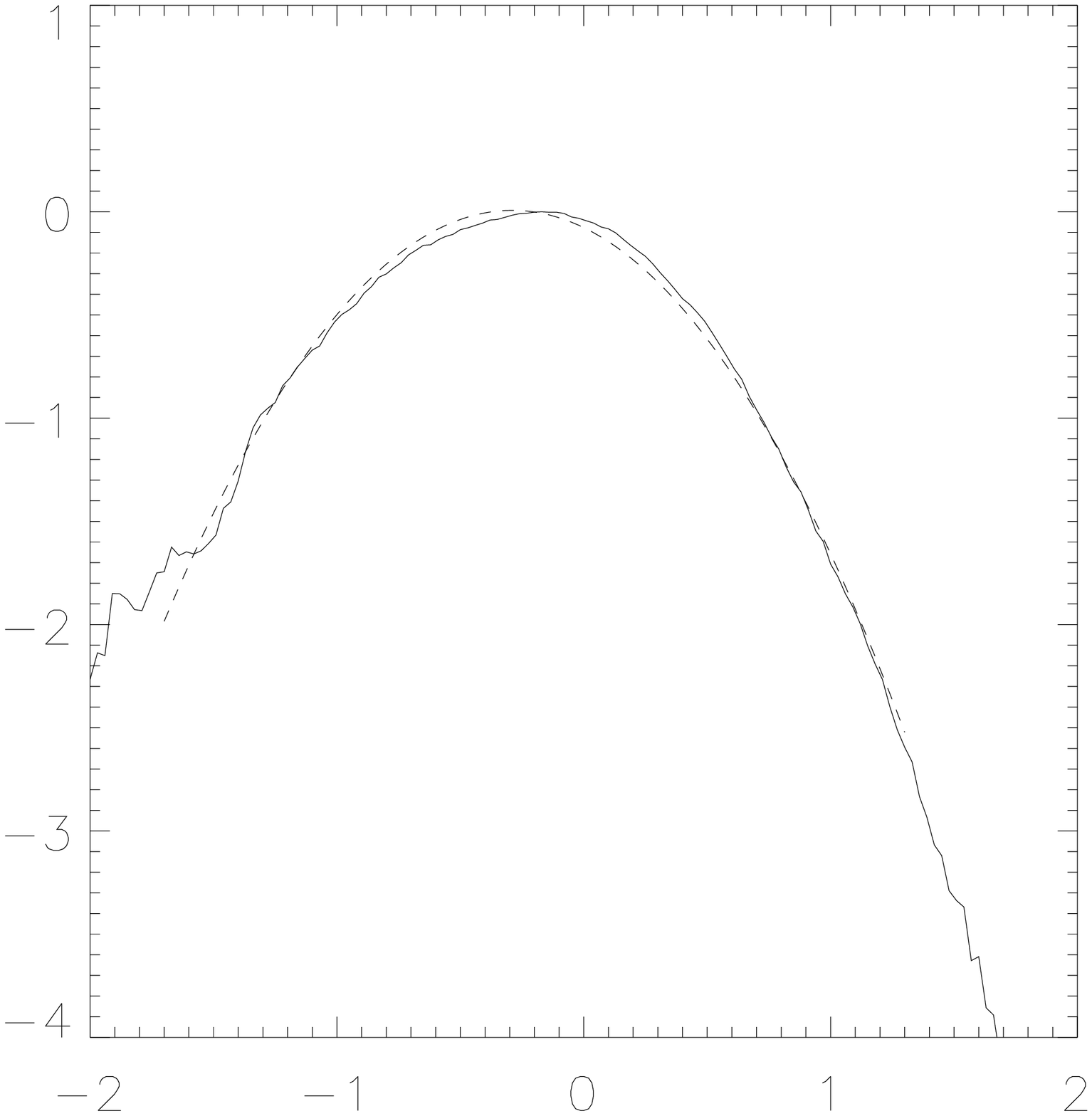,bbllx=.5cm,bblly=4cm,bburx=21.cm,bbury=24.5cm,height=7.cm,width=7.cm,clip=}}}
\caption{{\it Top}: Snapshot of $\log_{10}\rho$ (solid) and $\log_{10}|b|^2$
(dotted) for the run of Fig. 9. {\it Bottom}: Density PDF for the run of Fig. 9 in log-log
coordinates. The dashed shows a lognormal fit.}
\label{fig10}
\end{figure}

The behavior of the Alfv\'en wave pressure can also be tested numerically.
We have chosen to take $\beta=0$ (by turning off the thermal pressure term),
in order to make its effect more visible.
In a run with parallel propagation, the forcing is now applied on
modes 49-51 for the perpendicular velocity components and on modes
1-5 for the longitudinal component. Three different simulations have
been performed, with $\tilde M_a= 0.073, 0.795$ and $2.09$.
As visible on Figs. 11-13, while the magnetic pressure roughly
behaves as $\rho^{1/2}$ for $M_a= 0.4$, a clear tendency to approach
a polytropic behavior with $\gamma=2$ is observed when $M_a$ increases.
\begin{figure}[htb]
\centerline{\hbox{
\psfig{figure=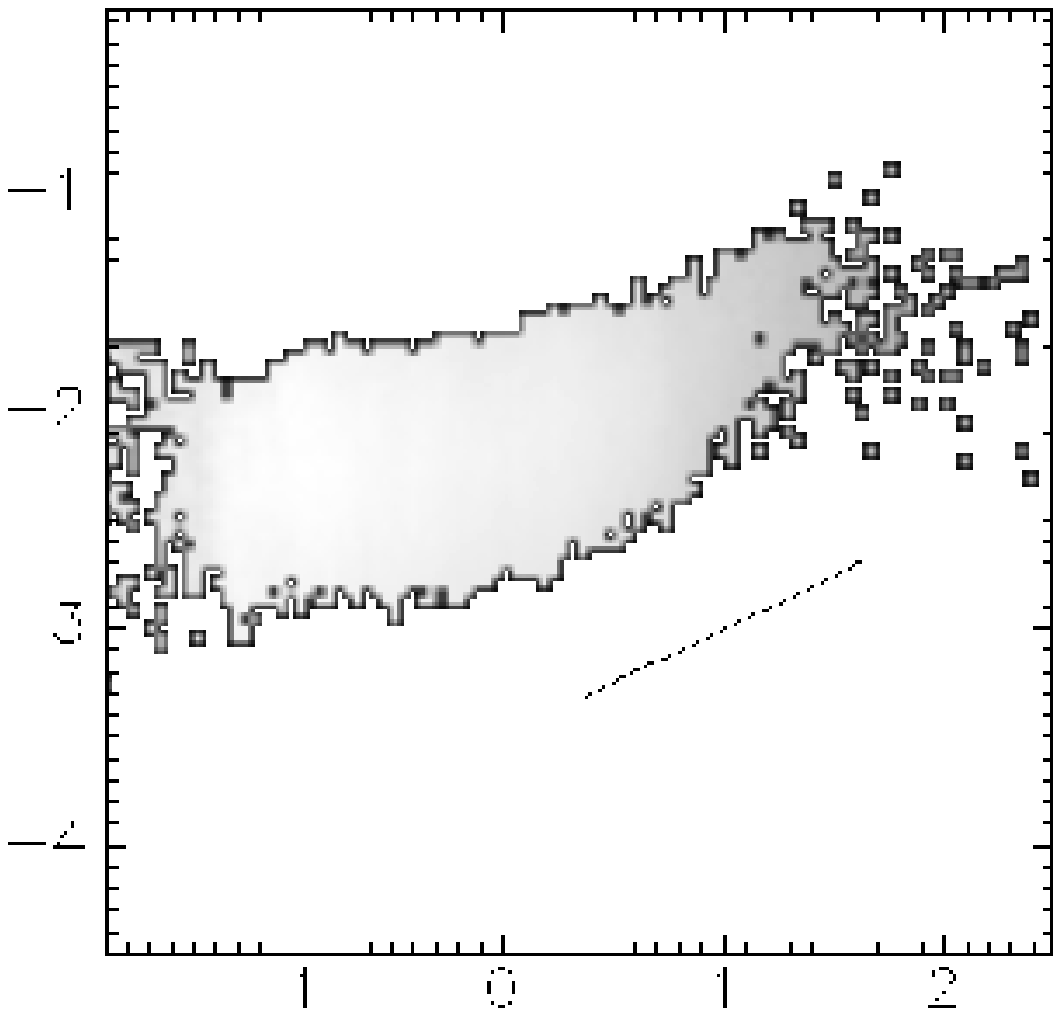,bbllx=4cm,bblly=8cm,bburx=16.cm,bbury=20.cm,height=8.cm,width=8.cm,clip=}}}
\caption{Log-log scatter plot of the magnetic pressure versus
density for a simulation with $\theta=0$ and $\beta=0$
The parameters are $\tilde M_a=0.073$,  ${\tilde
\frac{\delta\rho}{\rho}}=3.94$, ${\tilde \frac{\delta
B}{B}} =0.0564$ and $\tilde\sigma=0.60$. The line segment has a slope 0.5}
\label{fig11}
\end{figure}

\begin{figure}[htb]
\centerline{\hbox{
\psfig{figure=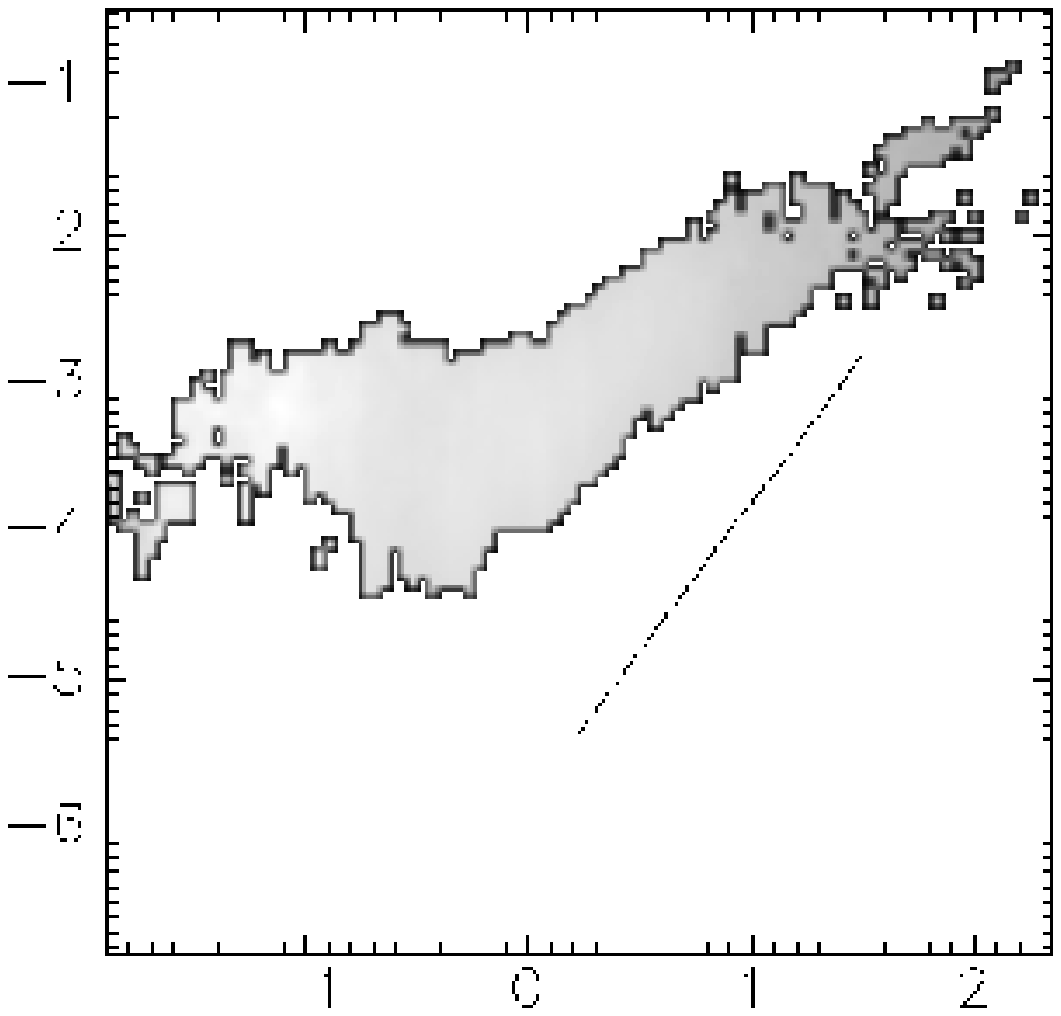,bbllx=4cm,bblly=8cm,bburx=16.cm,bbury=20.cm,height=8.cm,width=8.cm,clip=}}}
\caption{Log-log scatter plot of the magnetic pressure versus
density for a simulation with $\theta=0$ and $\beta=0$
The parameters are $\tilde M_a=0.795$,  ${\tilde
\frac{\delta\rho}{\rho}}=4.18$, ${\tilde \frac{\delta
B}{B}} =0.311$ and $\tilde\sigma=0.63$. The line segment has a slope 2.}
\label{fig12}
\end{figure}
\begin{figure}[htb]
\centerline{\hbox{
\psfig{figure=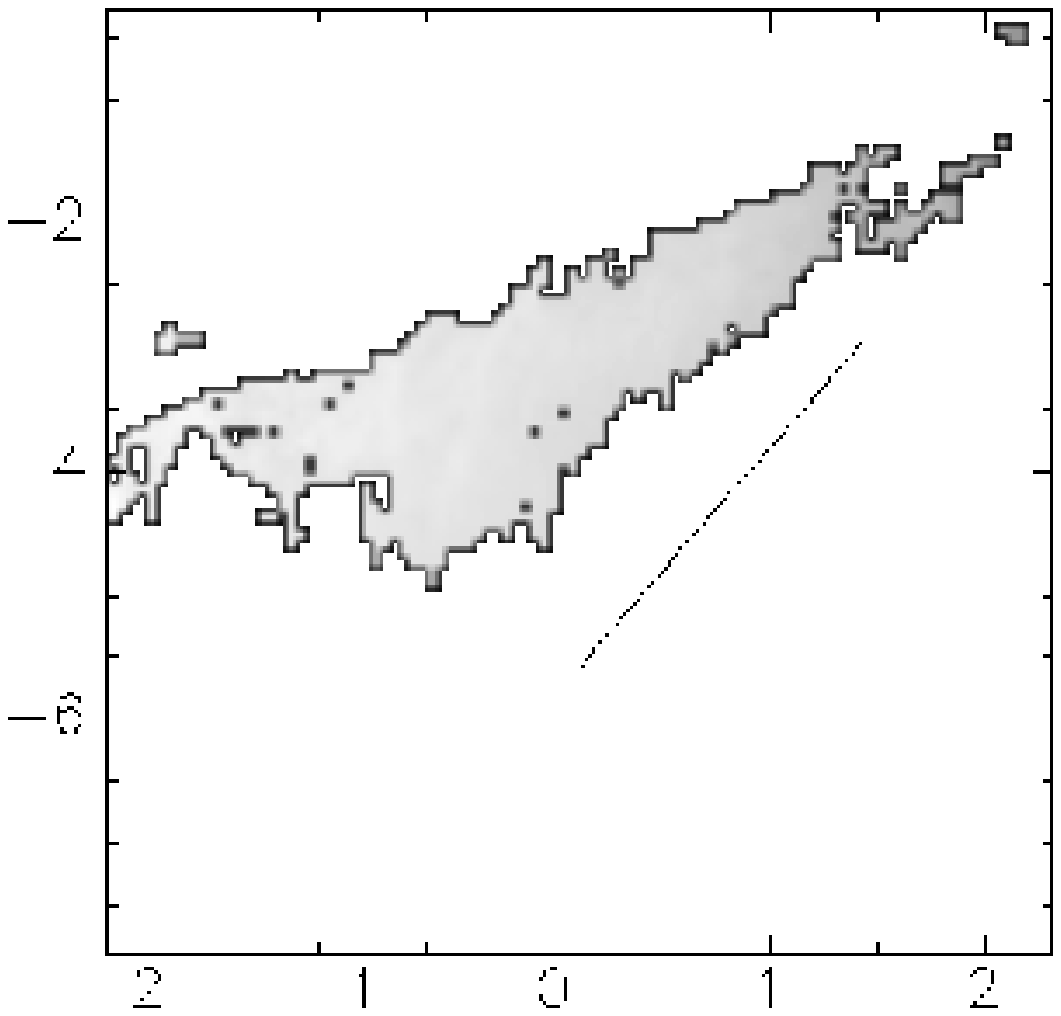,bbllx=4cm,bblly=8cm,bburx=16.cm,bbury=20.cm,height=8.cm,width=8.cm,clip=}}}
\caption{Log-log scatter plot of the magnetic pressure versus
density for a simulation with $\theta=0$ and $\beta=0$
The parameters are $\tilde M_a=2.09$,  ${\tilde
\frac{\delta\rho}{\rho}}=3.503$, ${\tilde \frac{\delta
B}{B}} =1.01$ and $\tilde\sigma=0.76$. The line segment has a slope 2.}
\label{fig13}
\end{figure}

%\section{Discussion and implications} \label{sec:discussion}

\section{Conclusions} \label{sec:conclusions}

\subsection{Summary and discussion} \label{sec:summary}

In this paper we have investigated the dependence of the magnetic
pressure, $B^2$ (and therefore, of the magnetic field strength) on
density in fully turbulent flows, parameterized by the Alfv\'enic Mach
number $M_a$ (which provides a more direct measure of the amount of
field distorsion than $\beta$) 
of the flow and the angle between the mean field direction and the
direction of wave propagation. To do this, we first employed the simple-wave
formalism to obtain insight by deriving relations between the
fluctuations of the magnetic and velocity fields on one hand, and the
density fluctuations on the other. We then presented numerical
experiments confirming the expectations from the analysis, and extended
the results.

From the simple-wave analysis, we concluded that the production of
density fluctuations is dominated by the slow mode at low $M_a$, while
at large $M_a$ both modes contribute. This result is in
disagreement with the assumption of Lithwick \& Goldreich (2001) that the
fast mode can be neglected in accounting for the spectrum of density
fluctuations because it decouples from the Alfv\'en and slow modes due
to its larger phase speed. Moreover, except for the case of small 
field fluctuations with simultaneously moderate density fluctuations and 
parallel field component, the magnetic pressure behaves as
\begin{equation}
P_{\rm mag}\simeq c_1 -c_2\rho
\end{equation}
for the slow mode, and as 
\begin{equation}
P_{\rm mag}\simeq \rho^2
\end{equation}
for the fast mode. This different scaling of the magnetic pressure with
density for the two modes is at the basis of the decorrelation observed, 
as in general a given value of the density can be arrived at by a
different kind of wave, and therefore have a different associated value
of the magnetic pressure. More generally, we can say that the
particular value of the magnetic pressure of a fluid parcel
will not be uniquely determined by its density, but instead, that it will
depend on the detailed history of how the density fluctuation was
arrived at. This also implies that {\it turbulent pressure cannot be simply
modeled with a polytropic law in general}, and in fact it can act as a
``random'' forcing instead of as a restoring force. In particular,
strong density fluctuations are possible even 
in the presence of a large uniform magnetic field, but in general
there is no relation between $\delta \rho/\rho$ and
$\tilde \beta$.

The angle $\theta$ between the the mean field and the direction of wave
propagation also plays an important role in determining the relative
importance of the modes. At perpendicular propagation, the slow mode
does not propagate, but at {\it almost} perpendicular propagation and
low $M_a$ it is dominant. At intermediate angles, the distinction
between the low- and high-$M_a$ cases is less pronounced, although at
large $M_a$ a tighter correlation between density and magnetic field
strength is observed at high densities. Also, in some cases (parallel 
propagation at moderate to large $M_a$), density peaks can even
correspond to magnetic pressure {\it minima}.  This is in fact
reminiscent of pressure balanced structures 
commonly observed in the solar wind and possibly present
in the ISM (see the simulations Mac Low et al 2001).

Our results have implications on the functional form of the density
PDF. We found that it tends to be lognormal, due to the unsystematic
action of the magnetic pressure, which allows the thermal pressure to
take control, except when the slow mode dominates density
fluctuation production. In this case, the strength of the random
forcing-like action of the magnetic pressure becomes density-dependent,
and causes a low-density excess in the PDF.

We also presented a perturbative analysis of the Alfv\'en-wave pressure, 
which recovered all the limiting polytropic cases obtained by
\cite{MZ95}, with $\gamma_e \simeq 2$ at large $M_a$, $\gamma_e
\simeq 3/2$ at moderate $M_a$, and $\gamma_e \simeq 1/2$ at low $M_a$,
but we also mentioned a result by Passot \& Gazol (in preparation) using 
this approach 
showing that the Alfv\'en wave pressure is in general not isotropic. As
pointed out by \cite{MZ95}, the conclusion of isotropy follows from
assuming a unique relation between the fluctuations of the velocity and
of the magnetic field, valid for a traveling Alfv\'en wave. That
isotropy does not hold in general is evidence that all
three wave modes contribute to the velocity fluctuations, each one with
a different functional form, similarly to the different scalings of magnetic
pressure with density for the slow and fast modes.

\subsection{Implications} \label{sec:implications}

Our results have a number of interesting implications on the standard
picture for the support of molecular clouds by magnetic fields, and
on the interpretation of observations. 
%
%In general, the standard picture
%assumes that cores can be supported against their self-gravity
%by magnetic fields if their ratio of mass to magnetic
%flux is subcritical (e.g., Mouschovias \& Spitzer 1976; Shu,
%Adams \& Lizano 1987), because it is assumed that flux freezing
%guarantees a positive scaling of field strength upon
%compression. However, this is a quasi-static picture, which moreover
%assumes a particular geometry. Instead, all numerical simulations to
%date suggest that core formation is a dynamic process (e.g., \VS, Passot \&
%Pouquet 1996; Klessen, Heitsch \& Mac Low 2000; Heitsch, Mac Low \&
%Klessen 2001; V\'azquez-Semadeni, Ballesteros-Paredes \& Klessen 2002),
%and analytical arguments can be given too (\VS, Shadmehri \&
%Ballesteros-Paredes 2002). In this case, 
%as we have seen here, different
%geometries (angles $\theta$) give rise to very different scalings of
%magnetic field strength with density.
%
Observational surveys of the magnetic field strength in molecular clouds 
seem to indicate that the former is essentially uncorrelated from the
density at low ($\lesssim 10^3$ cm$^{-3}$) densities, with recorrelation
seemingly appearing at higher densities (e.g., Crutcher et al.\
2002). In particular, in molecular clouds {\it non-detections of the
Zeeman effect are 
generally more frequent than detections}, and so in many cases only
upper limits to the field strength are available (e.g., Crutcher et
al. 1993; Bourke et al. 2001). Although it has been argued that these
results are consistent with the standard theory of molecular cloud
support (Shu, Adams \& Lizano 1987) through consideration of statistical
corrections to the random 
orientations of the field with respect to the line of sight (Crutcher et 
al. 1993; Crutcher et al. 2002), it is clear 
that they are also consistent with the lack of correlation between field 
strength and density found in the present paper and in other numerical
simulations (Padoan \& Nordlund 1999; Ostriker et al. 2001). This has
been already pointed out by Padoan \& Nordlund (1999).

The recorrelation apparently observed at higher densities (also  
noticeable in the 3D simulations of Ostriker et
al. (2001) including self-gravity) is consistent
with our cases with intermediate values of $\theta$ and large $M_a$,
suggesting that the Alfv\'enic Mach number in molecular clouds is
relatively large. However, it is also possible that this is an effect of 
self-gravity becoming important, which we have not considered here. It
is also possible that even at high densities the magnetic field strength 
is highly variable from one core to another. More 
observations of the magnetic field strength in high-density molecular
cloud cores, {\it reporting both detections and non detections}, are
needed.

%\noindent $\bullet$ $B^2$-$\rho$ correlation  determined by $M_a$
%and angle of propagation:\\
%\indent  $M_a\ll 1$: \BrickRed{slow} waves dominate: \\
%\centerline{\BrickRed{$P_{\rm mag}\simeq c_1 -c_2\rho$}}\\ 
%\indent contribution of \BrickRed{fast} waves: \\ \centerline{
%\BrickRed{$P_{\rm mag}\simeq \rho^2$}}
%
%\noindent $\bullet$ When fast waves are significant, slow waves are
%also present; large $B^2$-$\rho$ scatter.
%
%\noindent $\bullet$ Density fluctuations greatly depend on the
%dominant mode. Can be very \BrickRed{large even when ${\bf B}$ is strong} and 
%only slightly perturbed, but \BrickRed{no relation between $\frac{\delta
%\rho}{\rho}$ and $\beta$}.
%
%\noindent $\bullet$ Density PDF is {\Blue{log-normal due to $B^2$-$\rho$
%scatter} and shows \BrickRed{excess at small density when slow waves dominate}.
%
%\noindent $\bullet$ Magnetic pressure \PineGreen{cannot} be modeled as a
%polytropic law in general. Behaves as a forcing as well.

\begin{acknowledgements}
This work has received partial financial support from the French
national program PCMI (France) to T. P. and from CONACYT (M\'exico)
grant 27752-E to E. V.-S.
\end{acknowledgements}


\begin{thebibliography}{}

%\bibitem[1966]{baker} Baker, N. 1966, in Stellar Evolution, ed.\
%R. F. Stein,\& A. G. W. Cameron (Plenum, New York) 333
%
%\bibitem[1988]{balluch} Balluch, M. 1988, A\&A, 200, 58

%\bibitem[1999]{Allen_Shu99} Allen \& Shu 1999, 

\bibitem[Arons \& Max 1975]{AM75} Arons, J. \& Max, C. E. 1975, ApJ 196, 
L77

%\bibitem[1999]{BP_VS_S99} Balleteros-Paredes Hartmann \& VS 1999

%\bibitem[1993]{Blitz93} Blitz 1993

\bibitem[2001]{Bourke_etal01} Bourke, T. L., Myers, P. C., Robinson,
G., \& Hyland, A. R. 2001, ApJ 554, 916

\bibitem[2002]{CLV02} Cho, J., Lazarian, A. \& Vishniac, E. T. 2002, in
Simulations of Magnetohydrodynamic Turbulence in Astrophysics,
eds. E. Falgarone and T. Passot (Berlin, Springer Lecture Notes in
Physics), in press (astro-ph/0205286)

\bibitem[Cru_etal93]{Cru_etal93} Crutcher, R. M., Troland, T. H.,
Goodman, A. A., Heiles, C., Kazes, I., Myers, P. C. 1993, ApJ 407, 175

\bibitem[1999]{Crutcher99} Crutcher, R. M. 1999, ApJ 520, 706

\bibitem[1999]{CHT02} Crutcher, R., Heiles, C. \& Troland, T. 2002, in
Simulations of Magnetohydrodynamic Turbulence in Astrophysics,
eds. E. Falgarone and T. Passot (Berlin, Springer Lecture Notes in 
Physics), in press

\bibitem[1970]{Dewar70} Dewar, R. L. 1970, Phys. Fluids 13, 2710

\bibitem[1985]{Dickman85} Dickman, R. L. 1985, in Protostars and
Planets II, eds. D. C. Black and M. S. Matthews (Tucson: Univ. of
Arizona Press), 150

\bibitem{Ferraro 1955} Ferraro, V. C. A 1955, Proc. R. Soc. London, Ser. A,
223, 310

\bibitem[GO96]{GO96} Gammie, C. F. \& Ostriker, E. C. 1996, ApJ 466, 814

\bibitem[1993]{Heiles93} Heiles, C., Goodman, A. A., McKee, C. F. \&
Zweibel, E. G. , in Protostars and Planets III, eds. E. H. Levy and
J. I. Lunine (Tucson: Univ. of Arizona Press), 279

%\bibitem[HMK01]{HMK01} Heitsch, Mac Low \& Klessen 2001

\bibitem[HP00]{HP00} Hennebelle, P. \& P\'erault, M. 2000, A\&A 359, 1124

\bibitem[Kim et al. 2001]{KBM01} Kim, J., Balsara, D. \& Mac Low,
M.-M. 2001, JKAS 34, S333

%\bibitem[KHM00]{KHM00} Klessen, Heitsch \& Mac Low 2000

%\bibitem[2000]{Korn_Sca00} kornreich \& Scalo 2000

\bibitem[1987]{LL87} Landau, L. D. \& Lifshitz, E. M. 1987, Fluid
Mechanics (2nd ed.), Pergamon Press

%\bibitem[1981]{Larson81} Larson 1981

\bibitem[2001]{LG01} Lithwick, Y. \& Goldreich, P. 2001, ApJ 562, 279

\bibitem[1998]{MacLow_etal98} Mac Low, M.-M., Klessen, R. S., Burkert,
A. \& Smith, M. D. 1998, Phys. Rev. L. 80, 2754

\bibitem[2001]{MacLow_etal01} Mac Low, M.-M., Balsara, D., de Avillez,
M. A., Kim, J. 2001, astro-ph/0106509

\bibitem[Mann 1995]{Man95} Mann, G. 1995, J. Plasma Phys. 53, 109

\bibitem[MZ95]{MZ95} McKee, C.F. \& Zweibel, E.G. 1995, ApJ,
440, 686 (MZ95)

%\bibitem[MS76]{MS76} Mouschovias \& Spitzer 1976

\bibitem[1988]{Myers_Goodman88} Myers, P. C. \& Goodman, A. A. 1988,
ApJ, 326, L27

\bibitem[1999]{Nord_Pad99} Nordlund, \AA. \& Padoan. P. 1999, in
Interstellar Turbulence, eds. J. Franco and A. Carrami\~nana (Cambridge: 
Cambridge University Press), 218

\bibitem[2001]{OSG01} Ostriker, E. C., Stone, J, M. \& Gammie,
C. F. 2001, ApJ 546, 980

\bibitem[1999]{Pad_Nord99} Padoan , P.\& Nordlund, \AA. 1999, ApJ 526, 279

\bibitem[Passot \& \VS\ 1998]{PVS98} Passot, T. \& \VS, E. 1998,
Phys. Rev. E 58, 4501

%\bibitem[]{Passot \& Gazol, in preparation} 

\bibitem[1987]{Scalo 1987} Scalo, J. 1987, in Interstellar Processes,
eds. D. J. Hollenbach and H. A. Thronson (Dordrecht: Reidel), 349

\bibitem[SAL87]{SAL87} Shu, F., Adams, F. C. \& Lizano, S. 1987, ARAA 25, 23

\bibitem[1998]{Stone_etal98} Stone, J. M., Ostriker, E. C. \& Gammie, C.
F. 1998, ApJ 508, L99

%\bibitem[VPP96]{VPP96} V\'azquez-Semadeni, Passot \& Pouquet 1996 1998

\bibitem[1998]{VS_etal1998} V\'azquez-Semadeni, E., Cant\'o, J. \&
Lizano, S. 1998, ApJ 492, 596

\bibitem[2000]{VS_etal2000} V\'azquez-Semadeni, E., Ostriker, E. C.,
Passot, T., Gammie, C. F. \& Stone, J. M. 2000, in Protostars and
Planets IV, eds. V. Mannings, A. P. Boss \& S. Russell (Tucson:
University of Arizona Press), 3

%\bibitem[2002]{VBK02} V\'azquez-Semadeni, Ballesteros-Paredes \& Klessen 2002 

%\bibitem[VSB02]{VSB02} \VS, Shadmehri \& Ballesteros-Paredes 2002

\end{thebibliography}
\end{document}